\documentclass[aps,pra,reprint,superscriptaddress]{revtex4-2}

\usepackage{amssymb}
\usepackage{amsmath}
\usepackage[utf8]{inputenc}
\usepackage{textcomp}
\usepackage{subfigure}
\usepackage{graphicx}
\usepackage{bm} 
\usepackage{subcaption} 
\usepackage{theorem}
\usepackage{dsfont}
\newcommand{\idop}{\mathds{1}}

\newtheorem{definition}{Definition}
\newtheorem{proposition}{Proposition}

\newtheorem{theorem}[proposition]{Theorem}

\usepackage[ruled]{algorithm2e}
\usepackage{caption}
\newcommand{\bigtimes}{\raisebox{0.2ex}{\scalebox{1.5}{$\times$}}}

\newcommand{\Lin}{\mathrm{L}}

\DeclareMathOperator{\Tr}{Tr}  

\DeclareMathOperator{\rank}{rank}

\makeatletter
\newsavebox{\@brx}
\newcommand{\llangle}[1][]{\savebox{\@brx}{\ (\m@th{#1\langle}\)}%
  \mathopen{\copy\@brx\kern-0.5\wd\@brx\usebox{\@brx}}}
\newcommand{\rrangle}[1][]{\savebox{\@brx}{\ (\m@th{#1\rangle}\)}%
  \mathclose{\copy\@brx\kern-0.5\wd\@brx\usebox{\@brx}}}
\makeatother

\newcommand*{\cA}{\mathcal{A}}

\newcommand*{\cC}{\mathcal{C}}

\newcommand*{\cF}{\mathcal{F}}

\newcommand*{\cH}{\mathcal{H}}

\newcommand*{\cJ}{\mathcal{J}}

\newcommand*{\cM}{\mathcal{M}}

\newcommand*{\cT}{\mathcal{T}}

\newcommand{\bV}{\mathbb{V}}

\newcommand{\bS}{\mathbb{S}}
\newcommand{\bJ}{\mathbb{J}}

\newcommand{\tti}{\mathtt{i}}
\newcommand{\tto}{\mathtt{o}}
\newcommand{\fJ}{\mathfrak{J}}

\newcommand{\fS}{\mathfrak{S}}
\newcommand{\CC}{\mathbb{C}}

\newcommand{\st}{\mathrm{St}}

\begin{document}


\title{Efficient Self-Consistent Quantum Comb Tomography on the Product Stiefel Manifold}


\author{Xinlin He}%
\altaffiliation{These authors contributed equally to this work.}
\affiliation{State Key Lab of Millimeter Waves, Southeast University, Nanjing 211189, China}%
\affiliation{Frontiers Science Center for Mobile Information Communication and Security, Southeast University, Nanjing 210096, People's Republic of China}%

\author{Zetong Li}%
\altaffiliation{These authors contributed equally to this work.}
\affiliation{Thrust of Artificial Intelligence, Information Hub, The Hong Kong University of Science and Technology  (Guangzhou), Guangdong 511453, China}

\author{Congcong Zheng}%
\affiliation{State Key Lab of Millimeter Waves, Southeast University, Nanjing 211189, China}%
\affiliation{Frontiers Science Center for Mobile Information Communication and Security, Southeast University, Nanjing 210096, People's Republic of China}%

\author{Sixuan Li}%
\affiliation{State Key Lab of Millimeter Waves, Southeast University, Nanjing 211189, China}%
\affiliation{Frontiers Science Center for Mobile Information Communication and Security, Southeast University, Nanjing 210096, People's Republic of China}%

\author{Xutao Yu}%
\email{yuxutao@seu.edu.cn} 
\affiliation{State Key Lab of Millimeter Waves, Southeast University, Nanjing 211189, China}%
\affiliation{Frontiers Science Center for Mobile Information Communication and Security, Southeast University, Nanjing 210096, People's Republic of China}%

\author{Zaichen Zhang}%
\email{zczhang@seu.edu.cn} 
\affiliation{Frontiers Science Center for Mobile Information Communication and Security, Southeast University, Nanjing 210096, People's Republic of China}%
\affiliation{National Mobile Communications Research Laboratory, Southeast University, Nanjing 210096, China}


\date{\today}

\begin{abstract}
Characterizing non-Markovian quantum dynamics is currently hindered by the self-inconsistency and high computational complexity of existing quantum comb tomography (QCT)  methods. In this work, we propose a self-consistent framework that unifies the quantum comb, instrument set, and initial states into a single geometric entity, termed as the Comb-Instrument-State (CIS) set. We demonstrate that the CIS set naturally resides on a product Stiefel manifold, allowing the tomography problem to be solved via efficient unconstrained Riemannian optimization while automatically preserving physical constraints. Numerical simulations confirm that our approach is computationally scalable and robust against gate definition errors, significantly outperforming conventional isometry-based QCT methods. Our work indicates the potential to efficiently learn quantum comb with fewer computational resources.
\end{abstract}


\maketitle


\section{Introduction}

Quantum combs constitute a foundational model in quantum information and quantum computing, providing a powerful framework for describing complex quantum processes with finite quantum memory~\cite{wei2022Real,caruso2014Quantum,kos2023Circuits,maniscalco2007Entanglement,wolf2008Assessing,bylicka2014NonMarkovianity,paulson2021Hierarchy,zhao2022Quantum,sarovar2020Detecting,wei2022Hamiltonian,wilen2021Correlated,brownnutt2015Iontrap}. They formalize higher-order transformations as supermaps acting on quantum channels~\cite{wei2022Real,chiribella2009Theoretical, gutoski2007General, Bisio2011}. Owing to their strong universality and descriptive capacity, quantum combs have become indispensable for a wide range of tasks, including channel discrimination~\cite{PhysRevA.110.042210}, quantum metrology\cite{PhysRevLett.127.060501, Kurdziałek_2025}, quantum error correction~\cite{terhal2020Scalable,wang2023DGR}, and higher-order quantum computation~\cite{wei2022Real,caruso2014Quantum, chiribella2009Theoretical}. They also provide a natural mathematical framework for characterizing non-Markovian quantum noise, such as spatially and temporally correlated crosstalk\cite{zhao2022Quantum,sarovar2020Detecting,wei2022Hamiltonian, parrado-rodriguez2021Crosstalk,wang2023DGR}, which is ubiquitous in modern quantum devices~\cite{wilen2021Correlated,brownnutt2015Iontrap,papic2023Error, wei2022Hamiltonian,muller2019Understanding,terhal2020Scalable,kuhlmann2013Charge,yoneda2023Noisecorrelation,rojas-arias2023Spatial,harper2023Learning,suzuki2022Quantum}. Consequently, quantum comb tomography  (QCT)is of substantial importance and has emerged as a central challenge in quantum information science and quantum device engineering~\cite{chiribella2009Theoretical,milzQuantumStochasticProcesses2021a}.

Two fundamental challenges hinder the progress of QCT: \textit{self-inconsistency} and \textit{prohibitive complexity}. The former challenge stems from an implicit assumption in existing formulations that all non-target components of the experimental setup are known, echoing the core limitation of quantum process tomography  (QPT)~\cite{mohseni2008Quantumprocess, jkf7-wfcn}, which treats input states and measurements as fully characterized. However, this assumption always fails on real devices. Although current QCT techniques~\cite{white2020Demonstrationa,white2021Manybody,milzQuantumStochasticProcesses2021a, whiteNonMarkovianQuantumProcess2022, pollockNonMarkovianQuantumProcesses2018, milz2018Reconstructing} can reconstruct unknown combs when instruments or input states are known a priori, their performance degrades sharply once this knowledge is imperfect, leading to self-inconsistent reconstructions. 
The second challenge concerns computational complexity from two aspects: 
 (1) recent methods ~\cite{ milzQuantumStochasticProcesses2021a, whiteNonMarkovianQuantumProcess2022, pollockNonMarkovianQuantumProcesses2018} require a large number of parameters that scale with the product of squared input–output dimensions, exponentially increasing with time steps and system size~\cite{whiteNonMarkovianQuantumProcess2022}; 
 (2) The physical properties of a quantum comb require implementations of completely positive and causal  (CPC) constraints. This leads to introducing massive equality constraints to a positive semidefinite complex matrix~\cite{li2024non}.
These issues are amplified when self-consistency is pursued.

In this work, we propose a novel framework for self-consistent and computationally efficient QCT.  To address the self-inconsistency, recent efforts have attempted to jointly reconstruct quantum combs and instrument sets without prior state preparation, and measurement  (SPAM) calibration~\cite{li2024non, white2023Unifying}, which is inspired by gate set tomography  (GST)~\cite{greenbaum2015Introduction,nielsen2021Gate,  vinasMicroscopicParametrizationsGate2025}.  Building upon this direction, our key contribution is to identify and exploit the manifold geometry of the \emph{comb–instrument–state}  (CIS) set, which provides a unified and self-consistent description of the operational non-Markovian quantum processes. We show that all CIS components naturally reside on Stiefel manifolds, a special class of Riemannian manifolds, allowing the reconstruction task to be mapped onto the product Stiefel manifold. This geometric insight enables unconstrained Riemannian optimization, automatically enforcing physical constraints 
without the need for explicit projection or semidefinite programming, which sharply reduces the computational complexity. Manifold-based techniques have proven highly effective for constrained quantum learning~\cite{Abrudan2008, Wen2013, Zhu2025, PhysRevLett.130.150402, PhysRevLett.134.010803}, and our framework extends the principle in~\cite{PhysRevLett.134.010803} to substantially improve computational efficiency for this self-consistent QCT. We further analyze the computational complexity of the proposed method and validate its performance through numerical simulations.

\begin{figure*}
  \centering
  \includegraphics[width=1\linewidth]{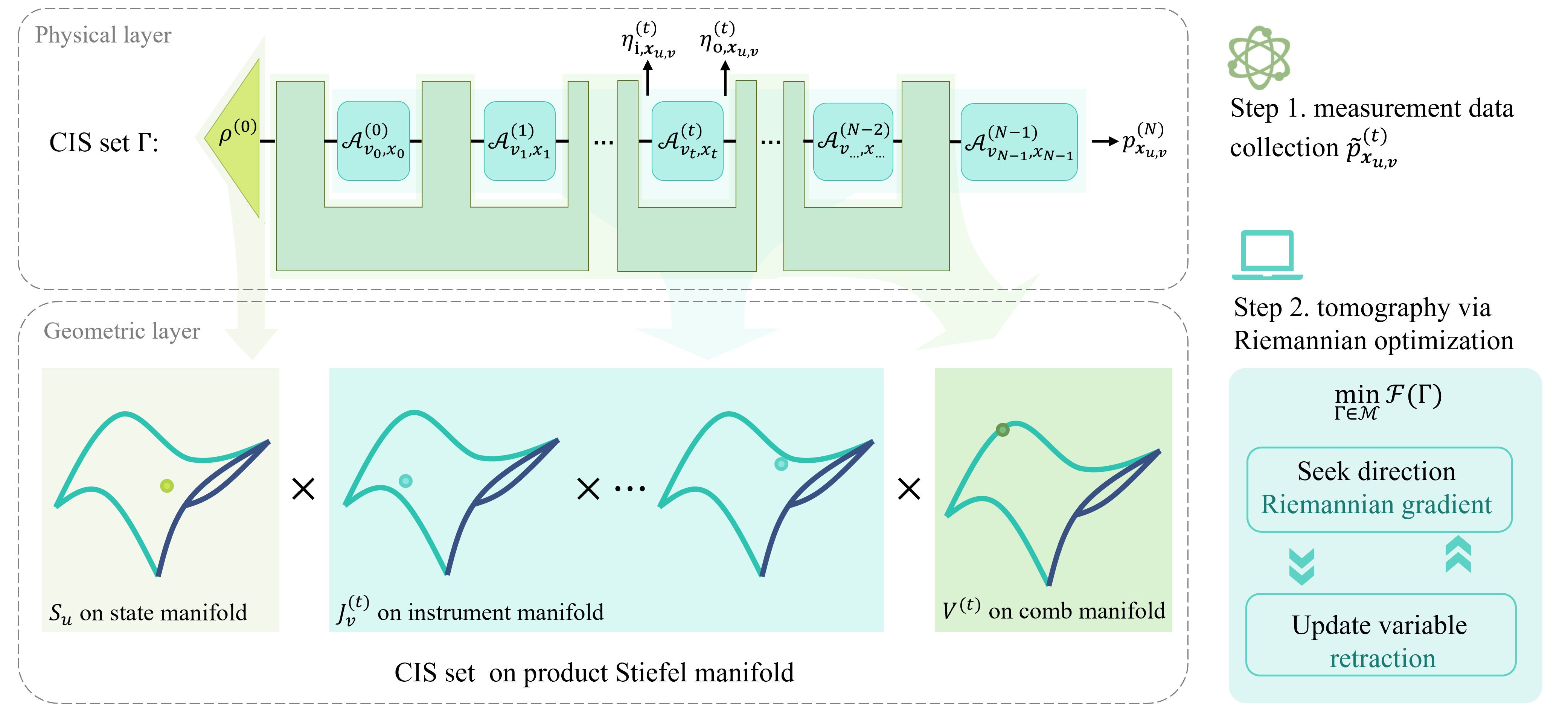}
  \captionsetup{justification=raggedright, singlelinecheck=false}   
  \caption{The general framework of self-consistent quantum comb tomography  (QCT). There is an operational open quantum process based on a quantum comb  (the green part). The initial state $\rho^{ (0)}$ may contain system–environment entanglement.  The maps $\cA^{ (t)}$ denote the experimentally accessible operations, and the comb governs the joint system–environment evolution.
  They jointly form a comb-instrument-state  (CIS) set, with certain geometric properties: every component of the CIS set is on a  (product) Stiefel manifold. The tomography framework contains 2 steps. After collecting the data from the quantum device, the second step is mainly based on the principle of manifold optimization.  $\cF (\Gamma)$ is the loss function, which is related to the intermediate state $\eta^{ (t)}_{\mathtt{i}}, \eta^{ (t)}_{\mathtt{o}}$ and the output probabilities $p^{ (t)}$ in the process. Finally, the core optimization steps are the calculation of the Riemannian gradient and retraction of the product manifold. }
  \label{IST_iQCT}
\end{figure*}

\section{Manifold Geometry of Comb-Instrument-State Set}

Fig.~\ref{IST_iQCT} gives a general picture for this work. 
Consider a comb with an initial state input and instrument interventions at all slots. It is an extension formalism of the standard quantum comb, and can be used to characterize the general supermaps of quantum processes with multiple input-output quantum operations.
For instance, the execution of a typical quantum device with Non-Markovian quantum noise is naturally described within this representation ~\cite{whiteNonMarkovianQuantumProcess2022}.

The $N$-time-step quantum comb $\cC^{ (N)}$ maps states from $\Lin (\cH_{\mathtt{i}})$ to $\Lin (\cH_{\mathtt{o}})$, where $\cH_{\tti} = \cH_{\tti_0}\otimes\dots\otimes \cH_{\tti_{N-1}}$ and $\cH_{\tto} = \cH_{\tto_0}\otimes \dots\otimes \cH_{\tto_{N-1}}$. An initial state $\rho^{{ (0)}}_{u}\in \Lin (\cH_{\mathtt{i}_0})$ from initial state set $\fS := \{\rho^{{ (0)}}_{u}\}$ is input at the beginning of the comb. At slot $t$, an instrument $\cJ^{ (t)}: (\cA^{ (t)}_1,\dots,\cA^{ (t)}_{|\cJ^{ (t)}|})$ is applied from an instrument set $\mathfrak{J}^{ (t)}:=\{\cJ^{ (t)}_v\}$, where $|\cJ^{ (t)}|$ represents the number of CP maps in $\cJ^{ (t)}$, $\cA^{ (t)}_{x}: \Lin (\cH_{\tto_{t}})\to\Lin (\cH_{\tti_{t+1}})$ is a complete positive  (CP) map with outcome $x$, and $\sum_{x=1}^{|\cJ^{ (t)}|} \cA^{ (t)}_{x}$ is complete positive and trace-preserving  (CPTP). We always use the notation $|\cdot|$ to represent the cardinality of the operand henceforth. 

Then, the probability of a sequence of outcomes $\bm{x}_{u,\bm{v}}$ raised by interactions between the comb $\cC^{ (N)}$, initial state set $\rho^{ (0)}_{u}$, and instruments $\cJ^{ (0)}_{v_0},\dots,\cJ^{ (N-1)}_{v_{N-1}}$ is given by
\begin{align}\label{eq:probability}
    p_{\bm{x}_{u,\bm{v}}}^{ (N)} = \Tr[\cC^{ (N)} (\cA_{v_{N-1},x_{N-1}}^{ (N-1)}\circ\dots\circ\cA_{v_0,x_0}^{ (0)}) (\rho^{ (0)}_{u})],
\end{align}
where the operational interpretation can be the probability that the supermap $\cC$ maps the CP map $\cA_{v_{N-1},x_{N-1}}^{ (N-1)}\circ\dots\circ\cA_{v_0,x_0}^{ (0)}$ to $\cC^{ (N)} (\cA_{v_{N-1},x_{N-1}}^{ (N-1)}\circ\dots\circ\cA_{v_0,x_0}^{ (0)})$, and then acts on the state $\rho^{ (0)}_{u}$.

In this scheme, the probability of any arbitrary sequence of outcomes is definite when the comb $\cC^{ (N)}$, the initial state set $\fS$, and the instrument sets $\fJ^{ (t)}$, $t=0,\dots, N-1$, are given. Thus, we call those parties that determine probabilities the comb-instrument-state  (CIS) set.

\begin{definition}[CIS set]
    Given an $N$-time step comb $\cC^{ (N)}:\Lin (\cH_{\mathtt{i}})\to\Lin (\cH_{\mathtt{o}})$, finite instrument sets $\mathfrak{J}^{ (t)}:=\{\cJ^{ (t)}_v:\Lin (\cH_{\tto_{t}})\to\Lin (\cH_{\tti_{t+1}})\}_{v=1}^{|\mathfrak{J}^{ (t)}|}$, $t=0,\dots,N-1$, and a finite initial state set $\fS:=\{\rho^{{ (0)}}_{u}\in\Lin (\cH_{\tti_{0}})\}_{u=1}^{|\fS|}$, the CIS set is defined as 
    \begin{align}
        \Gamma :=\Bigl\{\cC^{ (N)},\fJ:=\{\fJ^{ (t)}\}_{t=0}^{N-1},\fS\Bigr\}.
    \end{align}
\end{definition}

Optimizing a smooth function $\cF$ with respect to the CIS set generally involves solving the problem
\begin{align}
    \min_{\Gamma}~~~& \cF (\Gamma)\\
    s.t.~~~& \Upsilon^{ (N)}\ge 0,\label{cst:comb_choi_cp}\\
    &\Tr_{\tto_{t}} \Upsilon^{ (t+1)} = \idop_{\tti_{t}} \otimes \Upsilon^{ (t)}, ~~\forall t, ~~\Upsilon^{ (0)}=1,\label{cst:comb_choi_causal}\\
    &A^{ (t)}_{v,x}\ge 0, \forall t,v,x, \label{cst:map_choi_cp}\\
    &\sum_{x=1}^{|\cJ^{ (t)}|}\Tr_{\tti_{t+1}} A^{ (t)}_{v,x} = \idop_{\tto_{t}}, \forall t,v,\label{cst:map_choi_tp}\\
    &\rho^{ (0)}_u\ge 0, \Tr\rho^{ (0)}_u = 1,\label{cst:state_is_state}
\end{align}
where $\Upsilon^{ (N)}\in \Lin (\cH_{\mathtt{i}})\otimes\Lin (\cH_{\mathtt{o}})$ is the Choi state of $\cC^{ (N)}$ defined as
\begin{align}
    \cC^{{N}} (\rho_{\tti}\in \Lin (\cH_{\mathtt{i}})) = \Tr_{\tti}[\Upsilon^{ (N)} (\idop_{\tto}\otimes \rho_{\tti}^T)],
\end{align}
and $A^{ (t)}_{v,x}$ represent the Choi state of $\cA^{ (t)}_{v,x}$ given by 
\begin{align}
    \cA^{ (t)}_{x_i}[\rho\in\Lin (\cH_{\tto_{t}})) = \Tr_{\tto_t} (A^{ (t)}_{x} (\idop_{\tti_{t+1}}\otimes \rho^T)].
\end{align}
Constraint \eqref{cst:comb_choi_cp} states that the comb is CP, while \eqref{cst:comb_choi_causal} represents the causal property of the comb that the previous output system cannot be influenced by the future inputs. Constraints \eqref{cst:map_choi_cp} and \eqref{cst:map_choi_tp} limit that the instruments are CPTP. The initial state constraints are in \eqref{cst:state_is_state}.

Solving this problem is intractable when $\cF$ is non-convex, in which case the semi-definite programming techniques are not available. Moreover, when we focus on the comb with bounded quantum memory and instruments as well as the initial states with bounded environment dimensions, additional rank constraints are introduced to $\Upsilon^{ (t)}$, $A^{ (t)}_{u,x}$, and $\rho^{ (0)}_u$, for all $t,u,x,v$. Therefore, limited rank optimization for the CIS set is required in this situation, which is still open.

One of our main results states that the CIS set can be efficiently parameterized on the product Stiefel manifold, which enables the limited-rank unconstrained Riemannian optimization.

\begin{theorem}\label{thm:prod_st_man_rep}
    The CIS set $\Gamma =\{\cC^{ (N)},\fJ,\fS\}$ has the parameterization of 
    \begin{align}
        &\Gamma: \Gamma (\bV, \bJ, \bS),\\
        &\left\{\begin{aligned}
            &\bV:=  (V^{ (0)},\dots, V^{ (N-1)}),\\
            &V^{ (t)}\in \CC^{d_{\tti_{t}}d_{{a_t}}\times d_{\tto_{t}}d_{{a_{t+1}}}},\forall t
        \end{aligned}\right\}
        \\
        &\left\{\begin{aligned}
            &\bJ:= \{ (J^{ (t)}_1,\dots,J^{ (t)}_{|\fJ^{ (t)}|})\}_{t=0}^{N-1},\\
            &J^{ (t)}_{v}\in \CC^{\sum_{x=1}^{|\cJ^{ (t)}_{v}|}d_{e_{t,v,x}}\times d_{\tto_t}}, \forall t,v\\
        \end{aligned}\right\}\\
        &\left\{\begin{aligned}
            &\bS :=  (S_1,\dots, S_{|\fS|}), \\
            &S_{u}\in\CC^{d_{\tti_0}d_{r_u}\times 1}, \forall u
        \end{aligned}\right\},
    \end{align}
    where $d_{a_{t+1}}\ge \mathrm{rank} (\Upsilon^{ (t+1)})$, $d_{e_{t,v,x}}\ge \mathrm{rank} (A^{ (t)}_{v,x})$, $d_{r_{u}}\ge\mathrm{rank} (\rho^{ (0)}_u)$, with orthogonal constraints $V^{ (t)\dagger}V^{ (t)}=\idop_{d_{\tti_{t}}d_{{a_{t}}}}$, $J^{ (t)\dagger}_{v}J^{ (t)}_{v}=\idop_{d_{\tto_t}}$, and $S_u^\dagger S_u=1$.
    Then, $\Gamma (\bV, \bJ, \bS)$ has the product Stiefel manifold geometry
    \begin{equation}\label{eq:cis_prod_st}
        \begin{aligned}
            \cM := &\left[\bigtimes_{t=0}^{N-1} \st  (d_{\tti_{t}}d_{{a_t}},d_{\tto_{t}}d_{{a_{t+1}}}) \right]\\
            &\times\left[\bigtimes_{t=0}^{N-1}\bigtimes_{v=1}^{|\fJ^{ (t)}|} \st  (\sum_{x=1}^{|\cJ^{ (t)}_{v}|}d_{\tti_{t+1}}d_{e_{t,v,x}}, d_{\tto_t}) \right]\\
            &\times \left[\bigtimes_{u=1}^{|\fS|} \st  (d_{\tti_0}d_{r_u}, 1) \right],
        \end{aligned}
    \end{equation}
    where $\st (n,p):=\{X\in\CC^{n\times p}|X^\dagger X = \idop_{p}, n\ge p\}$ represents the complex Stiefel manifold.
\end{theorem}

The proof process is detailed in Appendix I.
Here, $\bV$ is derived from the isometry representation of the comb $\cC^{ (N)}$. That is, there exist isometries $V^{ (0)},\dots, V^{ (N-1)}$ that completely represent $\cC^{ (N)}$, such that
\begin{equation}
\begin{aligned}
    \cC^{ (N)}& (\rho_{\tti}\in\Lin (\cH_{\tti})) = \\
    &\Tr_{a_{N-1}}[V^{ (N-1)}\dots V^{ (0)}\rho_{\tti}V^{ (0)\dagger}\dots V^{ (N-1)^\dagger}],
\end{aligned}
\end{equation}
where $V^{ (t)}:\cH_{\tti_{t}}\otimes \cH_{a_{t}}\to \cH_{\tto_{t}}\otimes \cH_{a_{t+1}}$ satisfies $V^{ (t)\dagger}V^{ (t)}=\idop_{d_{\tto_{t}}d_{{a_{t+1}}}}$, and $\cH_{a_{t}}$ represents the $t$-th ancillary space with dimension $d_{a_t} \ge \mathrm{rank} (\Upsilon^{ (t)})$ corresponding to the bounded quantum memory at slot $t$, which indicates that $V^{ (t)}\in \st (d_{\tti_{t}}d_{{a_t}},d_{\tto_{t}}d_{{a_{t+1}}})$.

The parameterization of $\bJ$ results from the Stinespring dilation of instruments. Let $W^{ (t)}_{v,x}:\cH_{\tto_t}\to \cH_{\tti_{t+1}} \otimes \cH_{e_{t,v,x}}$ be the Stinespring dilation of $\cA^{ (t)}_{v,x}$ such that $\cA^{ (t)}_{v,x} (\cdot) = \Tr_{e_{t,v,x}}W^{ (t)\dagger}_{v,x} \cdot W^{ (t)}_{v,x}$ and $\sum_x W^{ (t)\dagger}_{v,x} W^{ (t)}_{v,x} = \idop_{d_{\tto_t}}$, where $\cH_{e_{t,v,x}}$ is the environment system with dimension $d_{e_{t,v,x}}\ge \mathrm{rank} (A^{ (t)}_{v,x})$. Then, we have $J_{v}^{ (t)}:=[W^{ (t)}_{v,1},\dots,W^{ (t)}_{v,|\cJ_{v}^{ (t)}|}]$ with orthogonal constraint $J_{v}^{ (t)^\dagger}J_{v}^{ (t)}=\idop_{d_{\tto_t}}$ that $J_{v}^{ (t)}\in\st (\sum_{x=1}^{|\cJ^{ (t)}_{v}|}d_{\tti_{t+1}}d_{e_{t,v,x}}, d_{\tto_t})$.

We note that the mixed state $\rho^{ (0)}_u\in \Lin (\cH_{\tti_0})$ can be purified as $S_u\in \cH_{\tti_0}\otimes \cH_{r_u}$ by introducing the $d_{r_u}\ge \mathrm{rank} (\rho^{ (0)}_u)$ reference system $\cH_{r_u}$, satisfying $S_u^\dagger S_u =1$ and $S_u\in\st (d_{\tti_0}d_{r_u}, 1)$. Then, we have \eqref{eq:cis_prod_st} by the product of the involved Stiefel manifolds together.

The Theorem~\ref{thm:prod_st_man_rep} enables the Riemannian optimization of the CIS set on the product Stiefel manifold.

\section{Tomography Framework based on Riemannian Optimization}\label{sec4}

\paragraph{Framework of self-consistent QCT} 
To capture general open-system behavior, the reconstruction of a complete CIS set is performed by collecting all experimentally observed outcome probabilities.  
As illustrated in Fig.~\ref{IST_iQCT}, the experimentally accessible components are quantum instruments.  
The experimenter applies a complete instrument set at state preparation and at every time slot, recording the results $\tilde{p}_{\bm{x}_{u,\bm{v}}}$, where $\bm{x}_{u,\bm{v}}$ corresponds to the sequence of outcomes in Eq.~\eqref{eq:probability}.  
Analogous to classical GST methods~\cite{greenbaum2015Introduction,nielsen2021Gate}, this framework defines a likelihood function that quantifies the discrepancy between model predictions and experimental data.  
Under the representation of Theorem~\ref{thm:prod_st_man_rep}, all observable statistics are uniquely determined by the CIS configuration described in the previous section.  
Consequently, the reconstruction task becomes the following unconstrained Riemannian optimization problem:
\begin{align}
    \min_{ (\bV,\bJ,\bS)\in\cM}\;
        \cF (\Gamma)
        =
        \sum_{t}\sum_{\bm{x}_{u,\bm{v}}} 
          \bigl|
            \tilde{p}^{ (t)}_{\bm{x}_{u,\bm{v}}}
            -
            p_{\bm{x}_{u,\bm{v}}}^{ (t)}
          \bigr|^{2}.
    \label{eq:ist_riemannian_opt}
\end{align}
Thus, CIS set tomography amounts precisely to estimating all manifold-valued variables from sequentially observed statistics.

Eq.~\eqref{eq:probability} forms the basis of the loss function $\cF$.  
It indicates the whole operational process with a series of intermediate states and the output probabilities.  
Since the causal structure of the comb guarantees that future instruments cannot influence past states or earlier comb isometries, the probabilities $p_{\bm{x}_{u,\bm{v}}}^{ (t)}$ at every slot admit the recursive representation associated with this formula. Specifically, as shown in the Fig.~\ref{IST_iQCT}, the intermediate states of this process at time $t$ satisfy
\begin{align}
    \eta^{ (t)}_{\mathtt{i},\bm{x}_{u,\bm{v}}}
    &= 
    V^{ (t)}
      \eta^{ (t-1)}_{\mathtt{o},\bm{x}_{u,\bm{v}}}
      V^{ (t)\dagger},\\[4pt]
    \eta^{ (t)}_{\mathtt{o},\bm{x}_{u,\bm{v}}}
    &=
    \Tr_{e_{t,v,x}}
      \Bigl[
          W_{v,x}^{ (t)}
          \eta^{ (t)}_{\mathtt{i},\bm{x}_{u,\bm{v}}}
          W_{v,x}^{ (t)\dagger}
      \Bigr],
\end{align}
where the outgoing state is generally unnormalized.

Since the CIS set resides on a product Stiefel manifold, we adopt the adaptive moment estimation  (ADAM) algorithm generalized to Riemannian geometry. It can be extended from Euclidean spaces to Riemannian manifolds by projecting Euclidean gradients onto tangent spaces and updating parameters via appropriate retractions~\cite{boumalIntroductionOptimizationSmooth2023a}, as detailed in Appendix II. This extension preserves the adaptive learning-rate and momentum mechanisms of standard ADAM~\cite{reddi2019convergenceadam}, while ensuring that each update respects the underlying manifold geometry. 
The algorithm is detailed in Appendix III. 

The optimization variables may be initialized either randomly or using prior experimental knowledge. When no prior information is available, a linear-inversion estimate can be computed, providing a high-quality starting point that improves convergence. Because all components of the CIS set lie on a product manifold, they can be updated jointly at each iteration, allowing the optimizer to converge toward a globally consistent solution. 
In addition, while we adopt ADAM for its stability and efficiency, other Riemannian optimization algorithms, such as conjugate-gradient or trust-region methods, are also applicable depending on the scale of the problem and convergence requirements. This flexibility ensures that the proposed framework can accommodate a wide range of CIS set tomography tasks.

\paragraph{Complexity} The primary computational cost arises from gradient evaluations during each iteration of the optimization process. 
Assume the input and output system dimensions are $d$. Let $T_t$ denote the number of iterations required for convergence at time step $t$, $d_{{e}}$ be the ancillary dimension of instruments, and let $n_A$ be the number of available instruments. The total computational complexity of estimating the full CIS set is then
\begin{align}\label{complexity}
\mathcal{O}\!\left ( \sum_{t=0}^{N-1} T_t O_t \right),~
O_t 
= \max\!\left\{
d^5 d_{{a}_t} d_{{a}_{t+1}}^3 d_{{e}}^3,
\;
n_A^3 d^3
\right\}.
\end{align}
The deviation process is detailed in Appendix IV. 
This expression accounts for the matrix multiplications and trace evaluations required at each gradient step.  
Compared with maximum-likelihood-based approaches, which require constrained optimization over high-dimensional parameter spaces, our framework is substantially more scalable.  
Furthermore, our method enables additional complexity reduction through dimensional compression of the ancillary Hilbert spaces in the quantum comb, guided by prior structural knowledge in ~\cite{PhysRevLett.134.010803}.  
This adaptive control over ancillary dimensions highlights the intrinsic scalability advantage of the Riemannian-geometry-based formulation.

\section{Numerical Experiment Results}

All simulations were conducted using the \textit{Quantum++} package~\cite{Gheorghiu2018} in C++. We address the task of non-Markovian gate set tomography. In our simulation, we generate a complete set of instruments and select a sequence of $N$ operations to act upon a prepared initial state. System-environment interactions and non-Markovian memory effects are introduced by coupling the system to a randomly generated quantum comb. The objective is to utilize collected measurement statistics to reconstruct the actual physical operations of the instrument set using our proposed framework. Informational completeness is secured by employing an overcomplete collection of completely positive trace-non-increasing  (CPTNI) instruments, comprising unitary operations implemented via ancillary systems and POVMs in the $X$, $Y$, and $Z$ bases. We performed simulations for single-qubit processes over two and three time steps.
\begin{figure}
\centering
\includegraphics[width=1\linewidth]{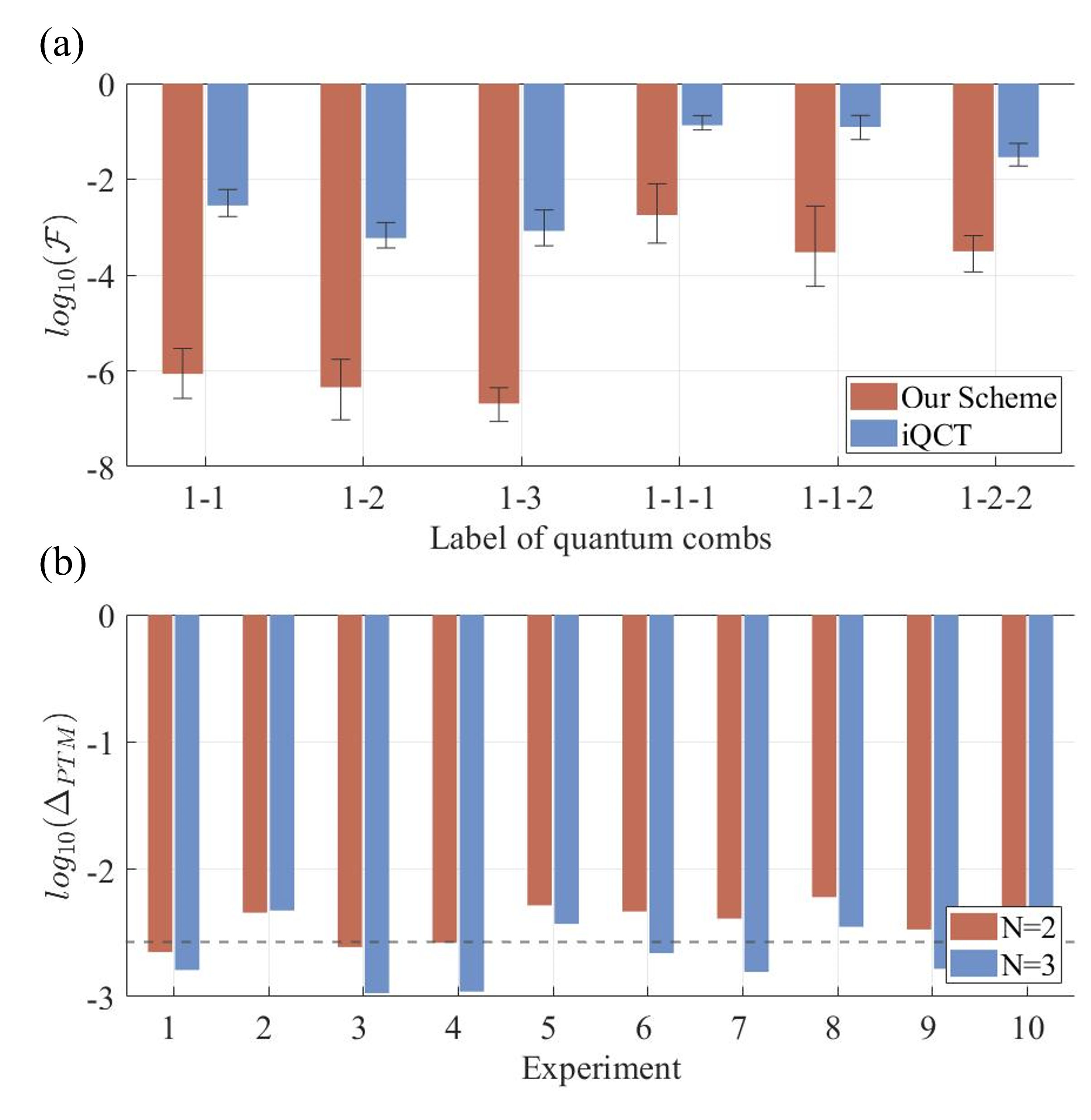}
\captionsetup{justification=raggedright, singlelinecheck=false}
\caption{ Numerical simulation results.
 (a) Comparison of loss values with the iQCT method. The bars show the average optimal loss under a certain quantum comb ancillary dimension. 
 (b) Difference of Pauli transfer matrix  (PTM) between practice and tomographic results. $\Delta_{PTM}$ is the average norm of the PTM difference of the whole instrument set.
}\label{cost}
\end{figure}

Fig.~\ref{cost} (a) compares the reconstruction performance across various ancillary dimensions; the label ``$n$--$m$'' indicates the dimension of the ancillary system at consecutive time steps. All optimization procedures were terminated when the gradient norm dropped below $10^{-5}$. We benchmark our approach against the isometry-based quantum comb tomography  (iQCT) method proposed in Ref.~\cite{PhysRevLett.134.010803}. To model device imperfections, we introduced small perturbations  (rotation angle 0.5 or 1) to the state preparation map, two CPTP maps, and one POVM element. In contrast to iQCT, which assumes perfect knowledge of local operations and consequently accumulates errors over time steps, our method demonstrates robustness by consistently maintaining lower reconstruction costs under identical perturbed conditions across various ancillary configurations for both two- and three-slot settings.

\begin{figure*}
\centering
\includegraphics[width=\linewidth]{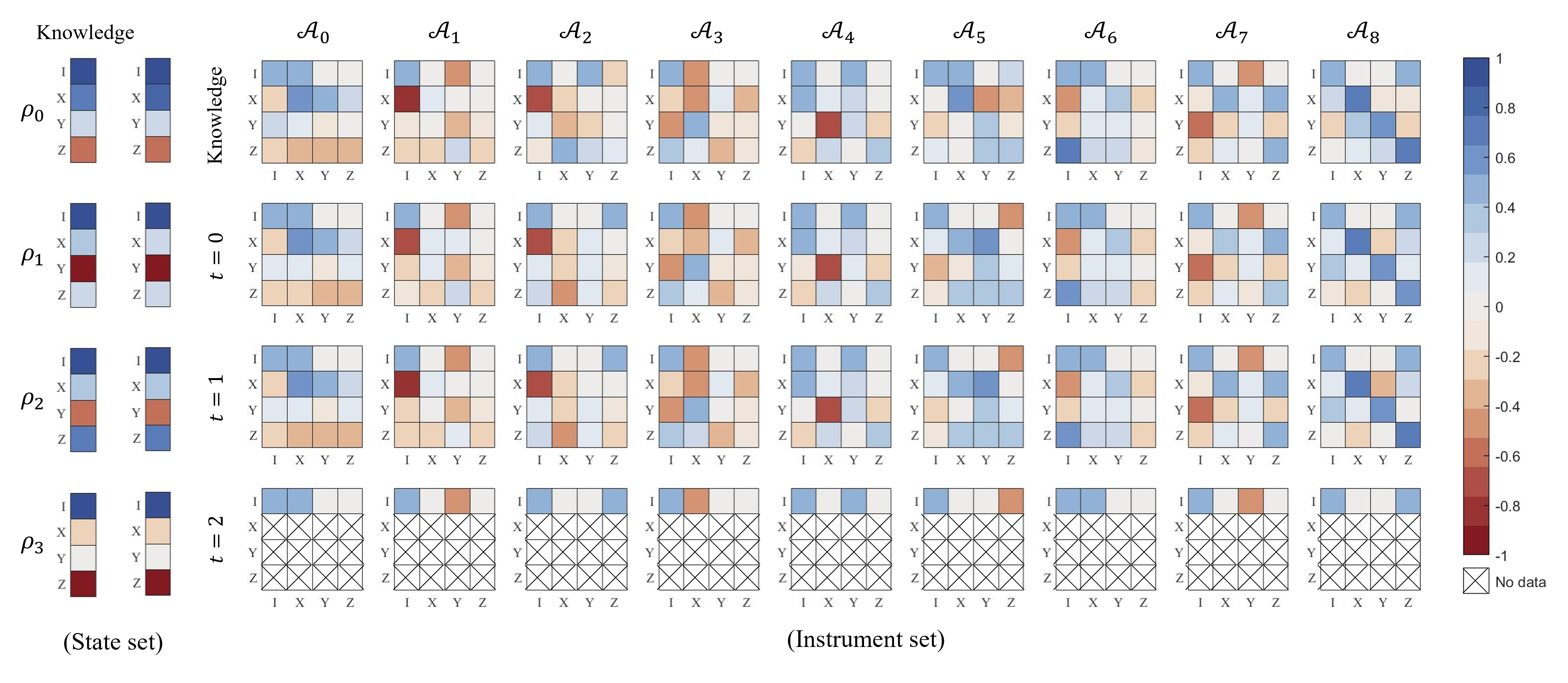}
\captionsetup{justification=raggedright, singlelinecheck=false}
\caption{Pauli transfer matrix  (PTM) tomography results in numerical simulations. There are the knowledge and tomographic results of the entire state set and the selected instrument set. For the left part, every $\rho_u$ is related to an initial state. For the right part, the first line is the knowledge, and the second line is the results at $t=0$, corresponding to the instrument's practice. The third and fourth lines are the same. }\label{tomo}
\end{figure*}

Fig.~\ref{tomo} displays representative reconstructed instruments in the Pauli transfer matrix  (PTM) representation. While the overcomplete set comprises 72 CPTNI instruments, we selected 9 representative examples for visualization. Among the displayed instruments, slight perturbations were introduced to $\cA_2$, $\cA_5$, and $\cA_8$ to simulate realistic experimental deviations. Consequently, the reconstructed maps for these instruments deviate slightly from the ideal prior models, accurately reflecting the actual physical operations. For the remaining unperturbed instruments, consistent recovery was achieved across all time steps despite the presence of environmental coupling and non-Markovian memory effects. It is important to note that at the final time step $t=2$, only the measurement operators can be reconstructed. The post-measurement states remain inherently inaccessible, which is a fundamental limitation of any tomography protocol.
In addition, Fig.~\ref{cost} (b) shows the norm of the PTM difference $\Delta_{PTM}$ between the tomographic results and practice, which contains the designed error of certain instruments. Here, the ancillary configuration of the quantum comb is "1-2-3". The low value of the norm directly reflects the great performance in this tomography task; our method can recover the quantum behavior under the non-Markovian effect.

\section{Conclusions}\label{sec6}
In this work, we develop a general, self-consistent, and computationally efficient framework for quantum comb tomography by exploiting the geometric structure of the CIS set. Through a unified parametrization of all CIS components on a product Stiefel manifold, we convert the conventionally constrained tomography task into an unconstrained Riemannian optimization problem, eliminating the need for explicit enforcement of complete positivity, trace conditions, and causal structure. This representation naturally captures bounded-memory effects and provides a geometric interpretation of system–environment interactions via isometry-based comb decompositions.
Built upon this manifold formulation, we introduce a smooth likelihood-minimization objective and implement a Stiefel-adapted Riemannian ADAM solver. Numerical simulations on non-Markovian gate set tomography demonstrate accurate and physically valid reconstructions under realistic noise, confirming that our approach effectively resolves the self-inconsistency issue in conventional QPT-based techniques while significantly reducing parameter overhead relative to generic Choi–Jamiołkowski models.

There are also many questions that remain open. Integrating experimental error mitigation and noise-aware models into the CIS tomography framework will enable more accurate reconstructions on real quantum hardware. Besides, extending the manifold-based representation to fully general quantum networks, including adaptive feedback and indefinite causal order, would further broaden its applicability. Finally, developing advanced Riemannian optimization algorithms and leveraging the geometric structure could improve convergence and scalability, opening opportunities for efficient characterization and control of complex quantum processes across larger systems and tasks.

\begin{acknowledgments}
This work was supported by the National Natural Science Foundation of China  (Grant No. 62471126), the Jiangsu Frontier Technology Research and Development Plan  (Grant No. BF2025066), and the Fundamental Research Funds for the Central Universities  (Grant No. 2242022k60001).
\end{acknowledgments}

\bibliography{trytry.bib}

\begin{thebibliography}{49}%
\makeatletter
\providecommand \@ifxundefined [1]{%
 \@ifx{#1\undefined}
}%
\providecommand \@ifnum [1]{%
 \ifnum #1\expandafter \@firstoftwo
 \else \expandafter \@secondoftwo
 \fi
}%
\providecommand \@ifx [1]{%
 \ifx #1\expandafter \@firstoftwo
 \else \expandafter \@secondoftwo
 \fi
}%
\providecommand \natexlab [1]{#1}%
\providecommand \enquote  [1]{``#1''}%
\providecommand \bibnamefont  [1]{#1}%
\providecommand \bibfnamefont [1]{#1}%
\providecommand \citenamefont [1]{#1}%
\providecommand \href@noop [0]{\@secondoftwo}%
\providecommand \href [0]{\begingroup \@sanitize@url \@href}%
\providecommand \@href[1]{\@@startlink{#1}\@@href}%
\providecommand \@@href[1]{\endgroup#1\@@endlink}%
\providecommand \@sanitize@url [0]{\catcode `\\12\catcode `\$12\catcode
  `\&12\catcode `\#12\catcode `\^12\catcode `\_12\catcode `\%12\relax}%
\providecommand \@@startlink[1]{}%
\providecommand \@@endlink[0]{}%
\providecommand \url  [0]{\begingroup\@sanitize@url \@url }%
\providecommand \@url [1]{\endgroup\@href {#1}{\urlprefix }}%
\providecommand \urlprefix  [0]{URL }%
\providecommand \Eprint [0]{\href }%
\providecommand \doibase [0]{https://doi.org/}%
\providecommand \selectlanguage [0]{\@gobble}%
\providecommand \bibinfo  [0]{\@secondoftwo}%
\providecommand \bibfield  [0]{\@secondoftwo}%
\providecommand \translation [1]{[#1]}%
\providecommand \BibitemOpen [0]{}%
\providecommand \bibitemStop [0]{}%
\providecommand \bibitemNoStop [0]{.\EOS\space}%
\providecommand \EOS [0]{\spacefactor3000\relax}%
\providecommand \BibitemShut  [1]{\csname bibitem#1\endcsname}%
\let\auto@bib@innerbib\@empty
\bibitem [{\citenamefont {Wei}\ \emph {et~al.}(2022{\natexlab{a}})\citenamefont
  {Wei}, \citenamefont {Jing}, \citenamefont {Zhang}, \citenamefont {Liao},
  \citenamefont {Yuan}, \citenamefont {Fan}, \citenamefont {Lyu}, \citenamefont
  {Zhou}, \citenamefont {Wang},\ and\ \citenamefont {Deng}}]{wei2022Real}%
  \BibitemOpen
  \bibfield  {author} {\bibinfo {author} {\bibfnamefont {S.-H.}\ \bibnamefont
  {Wei}}, \bibinfo {author} {\bibfnamefont {B.}~\bibnamefont {Jing}}, \bibinfo
  {author} {\bibfnamefont {X.-Y.}\ \bibnamefont {Zhang}}, \bibinfo {author}
  {\bibfnamefont {J.-Y.}\ \bibnamefont {Liao}}, \bibinfo {author}
  {\bibfnamefont {C.-Z.}\ \bibnamefont {Yuan}}, \bibinfo {author}
  {\bibfnamefont {B.-Y.}\ \bibnamefont {Fan}}, \bibinfo {author} {\bibfnamefont
  {C.}~\bibnamefont {Lyu}}, \bibinfo {author} {\bibfnamefont {D.-L.}\
  \bibnamefont {Zhou}}, \bibinfo {author} {\bibfnamefont {Y.}~\bibnamefont
  {Wang}},\ and\ \bibinfo {author} {\bibfnamefont {G.-W.}\ \bibnamefont
  {Deng}},\ }\bibfield  {title} {\bibinfo {title} {Towards real-world quantum
  networks: A review},\ }\href {https://doi.org/10.1002/lpor.202100219}
  {\bibfield  {journal} {\bibinfo  {journal} {Laser \& Photonics Reviews}\
  }\textbf {\bibinfo {volume} {16}},\ \bibinfo {pages} {2100219} (\bibinfo
  {year} {2022}{\natexlab{a}})}\BibitemShut {NoStop}%
\bibitem [{\citenamefont {Caruso}\ \emph {et~al.}(2014)\citenamefont {Caruso},
  \citenamefont {Giovannetti}, \citenamefont {Lupo},\ and\ \citenamefont
  {Mancini}}]{caruso2014Quantum}%
  \BibitemOpen
  \bibfield  {author} {\bibinfo {author} {\bibfnamefont {F.}~\bibnamefont
  {Caruso}}, \bibinfo {author} {\bibfnamefont {V.}~\bibnamefont {Giovannetti}},
  \bibinfo {author} {\bibfnamefont {C.}~\bibnamefont {Lupo}},\ and\ \bibinfo
  {author} {\bibfnamefont {S.}~\bibnamefont {Mancini}},\ }\bibfield  {title}
  {\bibinfo {title} {Quantum channels and memory effects},\ }\href
  {https://doi.org/10.1103/RevModPhys.86.1203} {\bibfield  {journal} {\bibinfo
  {journal} {Reviews of Modern Physics}\ }\textbf {\bibinfo {volume} {86}},\
  \bibinfo {pages} {1203} (\bibinfo {year} {2014})}\BibitemShut {NoStop}%
\bibitem [{\citenamefont {Kos}\ and\ \citenamefont
  {Styliaris}(2023)}]{kos2023Circuits}%
  \BibitemOpen
  \bibfield  {author} {\bibinfo {author} {\bibfnamefont {P.}~\bibnamefont
  {Kos}}\ and\ \bibinfo {author} {\bibfnamefont {G.}~\bibnamefont
  {Styliaris}},\ }\bibfield  {title} {\bibinfo {title} {Circuits of space and
  time quantum channels},\ }\href {https://doi.org/10.22331/q-2023-05-24-1020}
  {\bibfield  {journal} {\bibinfo  {journal} {Quantum}\ }\textbf {\bibinfo
  {volume} {7}},\ \bibinfo {pages} {1020} (\bibinfo {year} {2023})}\BibitemShut
  {NoStop}%
\bibitem [{\citenamefont {Maniscalco}\ \emph {et~al.}(2007)\citenamefont
  {Maniscalco}, \citenamefont {Olivares},\ and\ \citenamefont
  {Paris}}]{maniscalco2007Entanglement}%
  \BibitemOpen
  \bibfield  {author} {\bibinfo {author} {\bibfnamefont {S.}~\bibnamefont
  {Maniscalco}}, \bibinfo {author} {\bibfnamefont {S.}~\bibnamefont
  {Olivares}},\ and\ \bibinfo {author} {\bibfnamefont {M.~G.}\ \bibnamefont
  {Paris}},\ }\bibfield  {title} {\bibinfo {title} {Entanglement oscillations
  in non-{{Markovian}} quantum channels},\ }\href
  {https://doi.org/10.1103/PhysRevA.75.062119} {\bibfield  {journal} {\bibinfo
  {journal} {Physical Review A}\ }\textbf {\bibinfo {volume} {75}},\ \bibinfo
  {pages} {062119} (\bibinfo {year} {2007})}\BibitemShut {NoStop}%
\bibitem [{\citenamefont {Wolf}\ \emph {et~al.}(2008)\citenamefont {Wolf},
  \citenamefont {Eisert}, \citenamefont {Cubitt},\ and\ \citenamefont
  {Cirac}}]{wolf2008Assessing}%
  \BibitemOpen
  \bibfield  {author} {\bibinfo {author} {\bibfnamefont {M.~M.}\ \bibnamefont
  {Wolf}}, \bibinfo {author} {\bibfnamefont {J.}~\bibnamefont {Eisert}},
  \bibinfo {author} {\bibfnamefont {T.~S.}\ \bibnamefont {Cubitt}},\ and\
  \bibinfo {author} {\bibfnamefont {J.~I.}\ \bibnamefont {Cirac}},\ }\bibfield
  {title} {\bibinfo {title} {Assessing non-{{Markovian}} quantum dynamics},\
  }\href {https://doi.org/10.1103/PhysRevLett.101.150402} {\bibfield  {journal}
  {\bibinfo  {journal} {Physical review letters}\ }\textbf {\bibinfo {volume}
  {101}},\ \bibinfo {pages} {150402} (\bibinfo {year} {2008})}\BibitemShut
  {NoStop}%
\bibitem [{\citenamefont {Bylicka}\ \emph {et~al.}(2014)\citenamefont
  {Bylicka}, \citenamefont {Chru{\'s}ci{\'n}ski},\ and\ \citenamefont
  {Maniscalco}}]{bylicka2014NonMarkovianity}%
  \BibitemOpen
  \bibfield  {author} {\bibinfo {author} {\bibfnamefont {B.}~\bibnamefont
  {Bylicka}}, \bibinfo {author} {\bibfnamefont {D.}~\bibnamefont
  {Chru{\'s}ci{\'n}ski}},\ and\ \bibinfo {author} {\bibfnamefont
  {S.}~\bibnamefont {Maniscalco}},\ }\bibfield  {title} {\bibinfo {title}
  {Non-{{Markovianity}} and reservoir memory of quantum channels: A quantum
  information theory perspective},\ }\href {https://doi.org/10.1038/srep05720}
  {\bibfield  {journal} {\bibinfo  {journal} {Scientific reports}\ }\textbf
  {\bibinfo {volume} {4}},\ \bibinfo {pages} {5720} (\bibinfo {year}
  {2014})}\BibitemShut {NoStop}%
\bibitem [{\citenamefont {Paulson}\ \emph {et~al.}(2021)\citenamefont
  {Paulson}, \citenamefont {Panwar}, \citenamefont {Banerjee},\ and\
  \citenamefont {Srikanth}}]{paulson2021Hierarchy}%
  \BibitemOpen
  \bibfield  {author} {\bibinfo {author} {\bibfnamefont {K.~G.}\ \bibnamefont
  {Paulson}}, \bibinfo {author} {\bibfnamefont {E.}~\bibnamefont {Panwar}},
  \bibinfo {author} {\bibfnamefont {S.}~\bibnamefont {Banerjee}},\ and\
  \bibinfo {author} {\bibfnamefont {R.}~\bibnamefont {Srikanth}},\ }\bibfield
  {title} {\bibinfo {title} {Hierarchy of quantum correlations under
  non-{{Markovian}} dynamics},\ }\href
  {https://doi.org/10.1007/s11128-021-03061-9} {\bibfield  {journal} {\bibinfo
  {journal} {Quantum Information Processing}\ }\textbf {\bibinfo {volume}
  {20}},\ \bibinfo {pages} {1} (\bibinfo {year} {2021})}\BibitemShut {NoStop}%
\bibitem [{\citenamefont {Zhao}\ \emph {et~al.}(2022)\citenamefont {Zhao},
  \citenamefont {Linghu}, \citenamefont {Li}, \citenamefont {Xu}, \citenamefont
  {Wang}, \citenamefont {Xue}, \citenamefont {Jin},\ and\ \citenamefont
  {Yu}}]{zhao2022Quantum}%
  \BibitemOpen
  \bibfield  {author} {\bibinfo {author} {\bibfnamefont {P.}~\bibnamefont
  {Zhao}}, \bibinfo {author} {\bibfnamefont {K.}~\bibnamefont {Linghu}},
  \bibinfo {author} {\bibfnamefont {Z.}~\bibnamefont {Li}}, \bibinfo {author}
  {\bibfnamefont {P.}~\bibnamefont {Xu}}, \bibinfo {author} {\bibfnamefont
  {R.}~\bibnamefont {Wang}}, \bibinfo {author} {\bibfnamefont {G.}~\bibnamefont
  {Xue}}, \bibinfo {author} {\bibfnamefont {Y.}~\bibnamefont {Jin}},\ and\
  \bibinfo {author} {\bibfnamefont {H.}~\bibnamefont {Yu}},\ }\bibfield
  {title} {\bibinfo {title} {Quantum crosstalk analysis for simultaneous gate
  operations on superconducting qubits},\ }\href
  {https://doi.org/10.1103/PRXQuantum.3.020301} {\bibfield  {journal} {\bibinfo
   {journal} {PRX quantum}\ }\textbf {\bibinfo {volume} {3}},\ \bibinfo {pages}
  {020301} (\bibinfo {year} {2022})}\BibitemShut {NoStop}%
\bibitem [{\citenamefont {Sarovar}\ \emph {et~al.}(2020)\citenamefont
  {Sarovar}, \citenamefont {Proctor}, \citenamefont {Rudinger}, \citenamefont
  {Young}, \citenamefont {Nielsen},\ and\ \citenamefont
  {{Blume-Kohout}}}]{sarovar2020Detecting}%
  \BibitemOpen
  \bibfield  {author} {\bibinfo {author} {\bibfnamefont {M.}~\bibnamefont
  {Sarovar}}, \bibinfo {author} {\bibfnamefont {T.}~\bibnamefont {Proctor}},
  \bibinfo {author} {\bibfnamefont {K.}~\bibnamefont {Rudinger}}, \bibinfo
  {author} {\bibfnamefont {K.}~\bibnamefont {Young}}, \bibinfo {author}
  {\bibfnamefont {E.}~\bibnamefont {Nielsen}},\ and\ \bibinfo {author}
  {\bibfnamefont {R.}~\bibnamefont {{Blume-Kohout}}},\ }\bibfield  {title}
  {\bibinfo {title} {Detecting crosstalk errors in quantum information
  processors},\ }\href {https://doi.org/10.22331/q-2020-09-11-321} {\bibfield
  {journal} {\bibinfo  {journal} {Quantum}\ }\textbf {\bibinfo {volume} {4}},\
  \bibinfo {pages} {321} (\bibinfo {year} {2020})}\BibitemShut {NoStop}%
\bibitem [{\citenamefont {Wei}\ \emph {et~al.}(2022{\natexlab{b}})\citenamefont
  {Wei}, \citenamefont {Magesan}, \citenamefont {Lauer}, \citenamefont
  {Srinivasan}, \citenamefont {Bogorin}, \citenamefont {Carnevale},
  \citenamefont {Keefe}, \citenamefont {Kim}, \citenamefont {Klaus},\ and\
  \citenamefont {Landers}}]{wei2022Hamiltonian}%
  \BibitemOpen
  \bibfield  {author} {\bibinfo {author} {\bibfnamefont {K.~X.}\ \bibnamefont
  {Wei}}, \bibinfo {author} {\bibfnamefont {E.}~\bibnamefont {Magesan}},
  \bibinfo {author} {\bibfnamefont {I.}~\bibnamefont {Lauer}}, \bibinfo
  {author} {\bibfnamefont {S.}~\bibnamefont {Srinivasan}}, \bibinfo {author}
  {\bibfnamefont {D.~F.}\ \bibnamefont {Bogorin}}, \bibinfo {author}
  {\bibfnamefont {S.}~\bibnamefont {Carnevale}}, \bibinfo {author}
  {\bibfnamefont {G.~A.}\ \bibnamefont {Keefe}}, \bibinfo {author}
  {\bibfnamefont {Y.}~\bibnamefont {Kim}}, \bibinfo {author} {\bibfnamefont
  {D.}~\bibnamefont {Klaus}},\ and\ \bibinfo {author} {\bibfnamefont
  {W.}~\bibnamefont {Landers}},\ }\bibfield  {title} {\bibinfo {title}
  {Hamiltonian engineering with multicolor drives for fast entangling gates and
  quantum crosstalk cancellation},\ }\href
  {https://doi.org/10.1103/PhysRevLett.129.060501} {\bibfield  {journal}
  {\bibinfo  {journal} {Physical Review Letters}\ }\textbf {\bibinfo {volume}
  {129}},\ \bibinfo {pages} {060501} (\bibinfo {year}
  {2022}{\natexlab{b}})}\BibitemShut {NoStop}%
\bibitem [{\citenamefont {Wilen}\ \emph {et~al.}(2021)\citenamefont {Wilen},
  \citenamefont {Abdullah}, \citenamefont {Kurinsky}, \citenamefont {Stanford},
  \citenamefont {Cardani}, \citenamefont {{d'Imperio}}, \citenamefont {Tomei},
  \citenamefont {Faoro}, \citenamefont {Ioffe},\ and\ \citenamefont
  {Liu}}]{wilen2021Correlated}%
  \BibitemOpen
  \bibfield  {author} {\bibinfo {author} {\bibfnamefont {C.~D.}\ \bibnamefont
  {Wilen}}, \bibinfo {author} {\bibfnamefont {S.}~\bibnamefont {Abdullah}},
  \bibinfo {author} {\bibfnamefont {N.~A.}\ \bibnamefont {Kurinsky}}, \bibinfo
  {author} {\bibfnamefont {C.}~\bibnamefont {Stanford}}, \bibinfo {author}
  {\bibfnamefont {L.}~\bibnamefont {Cardani}}, \bibinfo {author} {\bibfnamefont
  {G.}~\bibnamefont {{d'Imperio}}}, \bibinfo {author} {\bibfnamefont
  {C.}~\bibnamefont {Tomei}}, \bibinfo {author} {\bibfnamefont
  {L.}~\bibnamefont {Faoro}}, \bibinfo {author} {\bibfnamefont {L.~B.}\
  \bibnamefont {Ioffe}},\ and\ \bibinfo {author} {\bibfnamefont {C.~H.}\
  \bibnamefont {Liu}},\ }\bibfield  {title} {\bibinfo {title} {Correlated
  charge noise and relaxation errors in superconducting qubits},\ }\href
  {https://doi.org/10.1038/s41586-021-03557-5} {\bibfield  {journal} {\bibinfo
  {journal} {Nature}\ }\textbf {\bibinfo {volume} {594}},\ \bibinfo {pages}
  {369} (\bibinfo {year} {2021})}\BibitemShut {NoStop}%
\bibitem [{\citenamefont {Brownnutt}\ \emph {et~al.}(2015)\citenamefont
  {Brownnutt}, \citenamefont {Kumph}, \citenamefont {Rabl},\ and\ \citenamefont
  {Blatt}}]{brownnutt2015Iontrap}%
  \BibitemOpen
  \bibfield  {author} {\bibinfo {author} {\bibfnamefont {M.}~\bibnamefont
  {Brownnutt}}, \bibinfo {author} {\bibfnamefont {M.}~\bibnamefont {Kumph}},
  \bibinfo {author} {\bibfnamefont {P.}~\bibnamefont {Rabl}},\ and\ \bibinfo
  {author} {\bibfnamefont {R.}~\bibnamefont {Blatt}},\ }\bibfield  {title}
  {\bibinfo {title} {Ion-trap measurements of electric-field noise near
  surfaces},\ }\href {https://doi.org/10.1103/RevModPhys.87.1419} {\bibfield
  {journal} {\bibinfo  {journal} {Reviews of modern Physics}\ }\textbf
  {\bibinfo {volume} {87}},\ \bibinfo {pages} {1419} (\bibinfo {year}
  {2015})}\BibitemShut {NoStop}%
\bibitem [{\citenamefont {Chiribella}\ \emph {et~al.}(2009)\citenamefont
  {Chiribella}, \citenamefont {D'Ariano},\ and\ \citenamefont
  {Perinotti}}]{chiribella2009Theoretical}%
  \BibitemOpen
  \bibfield  {author} {\bibinfo {author} {\bibfnamefont {G.}~\bibnamefont
  {Chiribella}}, \bibinfo {author} {\bibfnamefont {G.~M.}\ \bibnamefont
  {D'Ariano}},\ and\ \bibinfo {author} {\bibfnamefont {P.}~\bibnamefont
  {Perinotti}},\ }\bibfield  {title} {\bibinfo {title} {Theoretical framework
  for quantum networks},\ }\href {https://doi.org/10.1103/PhysRevA.80.022339}
  {\bibfield  {journal} {\bibinfo  {journal} {Phys. Rev. A}\ }\textbf {\bibinfo
  {volume} {80}},\ \bibinfo {pages} {022339} (\bibinfo {year}
  {2009})}\BibitemShut {NoStop}%
\bibitem [{\citenamefont {Gutoski}\ and\ \citenamefont
  {Watrous}(2007)}]{gutoski2007General}%
  \BibitemOpen
  \bibfield  {author} {\bibinfo {author} {\bibfnamefont {G.}~\bibnamefont
  {Gutoski}}\ and\ \bibinfo {author} {\bibfnamefont {J.}~\bibnamefont
  {Watrous}},\ }\bibfield  {title} {\bibinfo {title} {Toward a general theory
  of quantum games},\ }in\ \href {https://doi.org/10.1145/1250790.1250873}
  {\emph {\bibinfo {booktitle} {Proceedings of the Thirty-Ninth Annual ACM
  Symposium on Theory of Computing}}},\ \bibinfo {series and number} {STOC
  '07}\ (\bibinfo  {publisher} {Association for Computing Machinery},\ \bibinfo
  {address} {New York, NY, USA},\ \bibinfo {year} {2007})\ p.\ \bibinfo {pages}
  {565–574}\BibitemShut {NoStop}%
\bibitem [{\citenamefont {Bisio}\ \emph {et~al.}(2 24)\citenamefont {Bisio},
  \citenamefont {D’Ariano}, \citenamefont {Perinotti},\ and\ \citenamefont
  {Chiribella}}]{Bisio2011}%
  \BibitemOpen
  \bibfield  {author} {\bibinfo {author} {\bibfnamefont {A.}~\bibnamefont
  {Bisio}}, \bibinfo {author} {\bibfnamefont {G.~M.}\ \bibnamefont
  {D’Ariano}}, \bibinfo {author} {\bibfnamefont {P.}~\bibnamefont
  {Perinotti}},\ and\ \bibinfo {author} {\bibfnamefont {G.}~\bibnamefont
  {Chiribella}},\ }\bibfield  {title} {\bibinfo {title} {Minimal
  computational-space implementation of multiround quantum protocols},\ }\href
  {https://doi.org/10.1103/PhysRevA.83.022325} {\bibfield  {journal} {\bibinfo
  {journal} {Physical Review A}\ }\textbf {\bibinfo {volume} {83}},\ \bibinfo
  {pages} {022325} (\bibinfo {year} {2011-02-24})}\BibitemShut {NoStop}%
\bibitem [{\citenamefont {Zambon}(2024)}]{PhysRevA.110.042210}%
  \BibitemOpen
  \bibfield  {author} {\bibinfo {author} {\bibfnamefont {G.}~\bibnamefont
  {Zambon}},\ }\bibfield  {title} {\bibinfo {title} {Process tensor
  distinguishability measures},\ }\href
  {https://doi.org/10.1103/PhysRevA.110.042210} {\bibfield  {journal} {\bibinfo
   {journal} {Phys. Rev. A}\ }\textbf {\bibinfo {volume} {110}},\ \bibinfo
  {pages} {042210} (\bibinfo {year} {2024})}\BibitemShut {NoStop}%
\bibitem [{\citenamefont {Altherr}\ and\ \citenamefont
  {Yang}(2021)}]{PhysRevLett.127.060501}%
  \BibitemOpen
  \bibfield  {author} {\bibinfo {author} {\bibfnamefont {A.}~\bibnamefont
  {Altherr}}\ and\ \bibinfo {author} {\bibfnamefont {Y.}~\bibnamefont {Yang}},\
  }\bibfield  {title} {\bibinfo {title} {Quantum metrology for non-markovian
  processes},\ }\href {https://doi.org/10.1103/PhysRevLett.127.060501}
  {\bibfield  {journal} {\bibinfo  {journal} {Phys. Rev. Lett.}\ }\textbf
  {\bibinfo {volume} {127}},\ \bibinfo {pages} {060501} (\bibinfo {year}
  {2021})}\BibitemShut {NoStop}%
\bibitem [{\citenamefont {Kurdziałek}\ \emph {et~al.}(2025)\citenamefont
  {Kurdziałek}, \citenamefont {Dulian}, \citenamefont {Majsak}, \citenamefont
  {Chakraborty},\ and\ \citenamefont
  {Demkowicz-Dobrzański}}]{Kurdziałek_2025}%
  \BibitemOpen
  \bibfield  {author} {\bibinfo {author} {\bibfnamefont {S.}~\bibnamefont
  {Kurdziałek}}, \bibinfo {author} {\bibfnamefont {P.}~\bibnamefont {Dulian}},
  \bibinfo {author} {\bibfnamefont {J.}~\bibnamefont {Majsak}}, \bibinfo
  {author} {\bibfnamefont {S.}~\bibnamefont {Chakraborty}},\ and\ \bibinfo
  {author} {\bibfnamefont {R.}~\bibnamefont {Demkowicz-Dobrzański}},\
  }\bibfield  {title} {\bibinfo {title} {Quantum metrology using quantum combs
  and tensor network formalism},\ }\href
  {https://doi.org/10.1088/1367-2630/ada8d1} {\bibfield  {journal} {\bibinfo
  {journal} {New Journal of Physics}\ }\textbf {\bibinfo {volume} {27}},\
  \bibinfo {pages} {013019} (\bibinfo {year} {2025})}\BibitemShut {NoStop}%
\bibitem [{\citenamefont {Terhal}\ \emph {et~al.}(2020)\citenamefont {Terhal},
  \citenamefont {Conrad},\ and\ \citenamefont {Vuillot}}]{terhal2020Scalable}%
  \BibitemOpen
  \bibfield  {author} {\bibinfo {author} {\bibfnamefont {B.~M.}\ \bibnamefont
  {Terhal}}, \bibinfo {author} {\bibfnamefont {J.}~\bibnamefont {Conrad}},\
  and\ \bibinfo {author} {\bibfnamefont {C.}~\bibnamefont {Vuillot}},\
  }\bibfield  {title} {\bibinfo {title} {Towards scalable bosonic quantum error
  correction},\ }\href {https://doi.org/10.1088/2058-9565/ab98a5} {\bibfield
  {journal} {\bibinfo  {journal} {Quantum Science and Technology}\ }\textbf
  {\bibinfo {volume} {5}},\ \bibinfo {pages} {043001} (\bibinfo {year}
  {2020})}\BibitemShut {NoStop}%
\bibitem [{\citenamefont {Wang}\ \emph {et~al.}(2023)\citenamefont {Wang},
  \citenamefont {Liu}, \citenamefont {Liu}, \citenamefont {Gu}, \citenamefont
  {Baker}, \citenamefont {Chong},\ and\ \citenamefont {Han}}]{wang2023DGR}%
  \BibitemOpen
  \bibfield  {author} {\bibinfo {author} {\bibfnamefont {H.}~\bibnamefont
  {Wang}}, \bibinfo {author} {\bibfnamefont {P.}~\bibnamefont {Liu}}, \bibinfo
  {author} {\bibfnamefont {Y.}~\bibnamefont {Liu}}, \bibinfo {author}
  {\bibfnamefont {J.}~\bibnamefont {Gu}}, \bibinfo {author} {\bibfnamefont
  {J.}~\bibnamefont {Baker}}, \bibinfo {author} {\bibfnamefont {F.~T.}\
  \bibnamefont {Chong}},\ and\ \bibinfo {author} {\bibfnamefont
  {S.}~\bibnamefont {Han}},\ }\bibfield  {title} {\bibinfo {title} {{{DGR}}:
  {{Tackling Drifted}} and {{Correlated Noise}} in {{Quantum Error Correction}}
  via {{Decoding Graph Re-weighting}}},\ }\href@noop {} {\bibfield  {journal}
  {\bibinfo  {journal} {arXiv preprint arXiv:2311.16214}\ } (\bibinfo {year}
  {2023})},\ \Eprint {https://arxiv.org/abs/2311.16214} {arxiv:2311.16214}
  \BibitemShut {NoStop}%
\bibitem [{\citenamefont {{Parrado-Rodr{\'i}guez}}\ \emph
  {et~al.}(2021)\citenamefont {{Parrado-Rodr{\'i}guez}}, \citenamefont
  {{Ryan-Anderson}}, \citenamefont {Bermudez},\ and\ \citenamefont
  {M{\"u}ller}}]{parrado-rodriguez2021Crosstalk}%
  \BibitemOpen
  \bibfield  {author} {\bibinfo {author} {\bibfnamefont {P.}~\bibnamefont
  {{Parrado-Rodr{\'i}guez}}}, \bibinfo {author} {\bibfnamefont
  {C.}~\bibnamefont {{Ryan-Anderson}}}, \bibinfo {author} {\bibfnamefont
  {A.}~\bibnamefont {Bermudez}},\ and\ \bibinfo {author} {\bibfnamefont
  {M.}~\bibnamefont {M{\"u}ller}},\ }\bibfield  {title} {\bibinfo {title}
  {Crosstalk {{Suppression}} for {{Fault-tolerant Quantum Error Correction}}
  with {{Trapped Ions}}},\ }\href {https://doi.org/10.22331/q-2021-06-29-487}
  {\bibfield  {journal} {\bibinfo  {journal} {Quantum}\ }\textbf {\bibinfo
  {volume} {5}},\ \bibinfo {pages} {487} (\bibinfo {year} {2021})}\BibitemShut
  {NoStop}%
\bibitem [{\citenamefont {Papi{\v c}}\ \emph {et~al.}(2023)\citenamefont
  {Papi{\v c}}, \citenamefont {Auer},\ and\ \citenamefont {{de
  Vega}}}]{papic2023Error}%
  \BibitemOpen
  \bibfield  {author} {\bibinfo {author} {\bibfnamefont {M.}~\bibnamefont
  {Papi{\v c}}}, \bibinfo {author} {\bibfnamefont {A.}~\bibnamefont {Auer}},\
  and\ \bibinfo {author} {\bibfnamefont {I.}~\bibnamefont {{de Vega}}},\
  }\href@noop {} {\bibinfo {title} {Error {{Sources}} of {{Quantum Gates}} in
  {{Superconducting Qubits}}}} (\bibinfo {year} {2023}),\ \Eprint
  {https://arxiv.org/abs/2305.08916} {arxiv:2305.08916 [quant-ph]} \BibitemShut
  {NoStop}%
\bibitem [{\citenamefont {M{\"u}ller}\ \emph {et~al.}(2019)\citenamefont
  {M{\"u}ller}, \citenamefont {Cole},\ and\ \citenamefont
  {Lisenfeld}}]{muller2019Understanding}%
  \BibitemOpen
  \bibfield  {author} {\bibinfo {author} {\bibfnamefont {C.}~\bibnamefont
  {M{\"u}ller}}, \bibinfo {author} {\bibfnamefont {J.~H.}\ \bibnamefont
  {Cole}},\ and\ \bibinfo {author} {\bibfnamefont {J.}~\bibnamefont
  {Lisenfeld}},\ }\bibfield  {title} {\bibinfo {title} {Towards understanding
  two-level-systems in amorphous solids: Insights from quantum circuits},\
  }\href {https://doi.org/10.1088/1361-6633/ab3a7e} {\bibfield  {journal}
  {\bibinfo  {journal} {Reports on Progress in Physics}\ }\textbf {\bibinfo
  {volume} {82}},\ \bibinfo {pages} {124501} (\bibinfo {year}
  {2019})}\BibitemShut {NoStop}%
\bibitem [{\citenamefont {Kuhlmann}\ \emph {et~al.}(2013)\citenamefont
  {Kuhlmann}, \citenamefont {Houel}, \citenamefont {Ludwig}, \citenamefont
  {Greuter}, \citenamefont {Reuter}, \citenamefont {Wieck}, \citenamefont
  {Poggio},\ and\ \citenamefont {Warburton}}]{kuhlmann2013Charge}%
  \BibitemOpen
  \bibfield  {author} {\bibinfo {author} {\bibfnamefont {A.~V.}\ \bibnamefont
  {Kuhlmann}}, \bibinfo {author} {\bibfnamefont {J.}~\bibnamefont {Houel}},
  \bibinfo {author} {\bibfnamefont {A.}~\bibnamefont {Ludwig}}, \bibinfo
  {author} {\bibfnamefont {L.}~\bibnamefont {Greuter}}, \bibinfo {author}
  {\bibfnamefont {D.}~\bibnamefont {Reuter}}, \bibinfo {author} {\bibfnamefont
  {A.~D.}\ \bibnamefont {Wieck}}, \bibinfo {author} {\bibfnamefont
  {M.}~\bibnamefont {Poggio}},\ and\ \bibinfo {author} {\bibfnamefont {R.~J.}\
  \bibnamefont {Warburton}},\ }\bibfield  {title} {\bibinfo {title} {Charge
  noise and spin noise in a semiconductor quantum device},\ }\href
  {https://doi.org/10.1038/nphys2688} {\bibfield  {journal} {\bibinfo
  {journal} {Nature Physics}\ }\textbf {\bibinfo {volume} {9}},\ \bibinfo
  {pages} {570} (\bibinfo {year} {2013})}\BibitemShut {NoStop}%
\bibitem [{\citenamefont {Yoneda}\ \emph {et~al.}(2023)\citenamefont {Yoneda},
  \citenamefont {{Rojas-Arias}}, \citenamefont {Stano}, \citenamefont {Takeda},
  \citenamefont {Noiri}, \citenamefont {Nakajima}, \citenamefont {Loss},\ and\
  \citenamefont {Tarucha}}]{yoneda2023Noisecorrelation}%
  \BibitemOpen
  \bibfield  {author} {\bibinfo {author} {\bibfnamefont {J.}~\bibnamefont
  {Yoneda}}, \bibinfo {author} {\bibfnamefont {J.~S.}\ \bibnamefont
  {{Rojas-Arias}}}, \bibinfo {author} {\bibfnamefont {P.}~\bibnamefont
  {Stano}}, \bibinfo {author} {\bibfnamefont {K.}~\bibnamefont {Takeda}},
  \bibinfo {author} {\bibfnamefont {A.}~\bibnamefont {Noiri}}, \bibinfo
  {author} {\bibfnamefont {T.}~\bibnamefont {Nakajima}}, \bibinfo {author}
  {\bibfnamefont {D.}~\bibnamefont {Loss}},\ and\ \bibinfo {author}
  {\bibfnamefont {S.}~\bibnamefont {Tarucha}},\ }\bibfield  {title} {\bibinfo
  {title} {Noise-correlation spectrum for a pair of spin qubits in silicon},\
  }\href {https://doi.org/10.1038/s41567-023-02238-6} {\bibfield  {journal}
  {\bibinfo  {journal} {Nature Physics}\ }\textbf {\bibinfo {volume} {19}},\
  \bibinfo {pages} {1793} (\bibinfo {year} {2023})}\BibitemShut {NoStop}%
\bibitem [{\citenamefont {{Rojas-Arias}}\ \emph {et~al.}(2023)\citenamefont
  {{Rojas-Arias}}, \citenamefont {Noiri}, \citenamefont {Stano}, \citenamefont
  {Nakajima}, \citenamefont {Yoneda}, \citenamefont {Takeda}, \citenamefont
  {Kobayashi}, \citenamefont {Sammak}, \citenamefont {Scappucci},\ and\
  \citenamefont {Loss}}]{rojas-arias2023Spatial}%
  \BibitemOpen
  \bibfield  {author} {\bibinfo {author} {\bibfnamefont {J.~S.}\ \bibnamefont
  {{Rojas-Arias}}}, \bibinfo {author} {\bibfnamefont {A.}~\bibnamefont
  {Noiri}}, \bibinfo {author} {\bibfnamefont {P.}~\bibnamefont {Stano}},
  \bibinfo {author} {\bibfnamefont {T.}~\bibnamefont {Nakajima}}, \bibinfo
  {author} {\bibfnamefont {J.}~\bibnamefont {Yoneda}}, \bibinfo {author}
  {\bibfnamefont {K.}~\bibnamefont {Takeda}}, \bibinfo {author} {\bibfnamefont
  {T.}~\bibnamefont {Kobayashi}}, \bibinfo {author} {\bibfnamefont
  {A.}~\bibnamefont {Sammak}}, \bibinfo {author} {\bibfnamefont
  {G.}~\bibnamefont {Scappucci}},\ and\ \bibinfo {author} {\bibfnamefont
  {D.}~\bibnamefont {Loss}},\ }\bibfield  {title} {\bibinfo {title} {Spatial
  noise correlations beyond nearest neighbors in 28 {{Si}}/{{Si-Ge}} spin
  qubits},\ }\href {https://doi.org/10.1103/PhysRevApplied.20.054024}
  {\bibfield  {journal} {\bibinfo  {journal} {Physical Review Applied}\
  }\textbf {\bibinfo {volume} {20}},\ \bibinfo {pages} {054024} (\bibinfo
  {year} {2023})}\BibitemShut {NoStop}%
\bibitem [{\citenamefont {Harper}\ and\ \citenamefont
  {Flammia}(2023)}]{harper2023Learning}%
  \BibitemOpen
  \bibfield  {author} {\bibinfo {author} {\bibfnamefont {R.}~\bibnamefont
  {Harper}}\ and\ \bibinfo {author} {\bibfnamefont {S.~T.}\ \bibnamefont
  {Flammia}},\ }\bibfield  {title} {\bibinfo {title} {Learning correlated noise
  in a 39-qubit quantum processor},\ }\href
  {https://doi.org/10.1103/PRXQuantum.4.040311} {\bibfield  {journal} {\bibinfo
   {journal} {PRX Quantum}\ }\textbf {\bibinfo {volume} {4}},\ \bibinfo {pages}
  {040311} (\bibinfo {year} {2023})}\BibitemShut {NoStop}%
\bibitem [{\citenamefont {Suzuki}\ \emph {et~al.}(2022)\citenamefont {Suzuki},
  \citenamefont {Endo}, \citenamefont {Fujii},\ and\ \citenamefont
  {Tokunaga}}]{suzuki2022Quantum}%
  \BibitemOpen
  \bibfield  {author} {\bibinfo {author} {\bibfnamefont {Y.}~\bibnamefont
  {Suzuki}}, \bibinfo {author} {\bibfnamefont {S.}~\bibnamefont {Endo}},
  \bibinfo {author} {\bibfnamefont {K.}~\bibnamefont {Fujii}},\ and\ \bibinfo
  {author} {\bibfnamefont {Y.}~\bibnamefont {Tokunaga}},\ }\bibfield  {title}
  {\bibinfo {title} {Quantum error mitigation as a universal error reduction
  technique: {{Applications}} from the nisq to the fault-tolerant quantum
  computing eras},\ }\href {https://doi.org/10.1103/PRXQuantum.3.010345}
  {\bibfield  {journal} {\bibinfo  {journal} {PRX Quantum}\ }\textbf {\bibinfo
  {volume} {3}},\ \bibinfo {pages} {010345} (\bibinfo {year}
  {2022})}\BibitemShut {NoStop}%
\bibitem [{\citenamefont {Milz}\ and\ \citenamefont {Modi}(7
  14)}]{milzQuantumStochasticProcesses2021a}%
  \BibitemOpen
  \bibfield  {author} {\bibinfo {author} {\bibfnamefont {S.}~\bibnamefont
  {Milz}}\ and\ \bibinfo {author} {\bibfnamefont {K.}~\bibnamefont {Modi}},\
  }\bibfield  {title} {\bibinfo {title} {Quantum {{Stochastic Processes}} and
  {{Quantum}} non-{{Markovian Phenomena}}},\ }\href
  {https://doi.org/10.1103/PRXQuantum.2.030201} {\bibfield  {journal} {\bibinfo
   {journal} {PRX Quantum}\ }\textbf {\bibinfo {volume} {2}},\ \bibinfo {pages}
  {030201} (\bibinfo {year} {2021-07-14})}\BibitemShut {NoStop}%
\bibitem [{\citenamefont {Mohseni}\ \emph {et~al.}(2008)\citenamefont
  {Mohseni}, \citenamefont {Rezakhani},\ and\ \citenamefont
  {Lidar}}]{mohseni2008Quantumprocess}%
  \BibitemOpen
  \bibfield  {author} {\bibinfo {author} {\bibfnamefont {M.}~\bibnamefont
  {Mohseni}}, \bibinfo {author} {\bibfnamefont {A.~T.}\ \bibnamefont
  {Rezakhani}},\ and\ \bibinfo {author} {\bibfnamefont {D.~A.}\ \bibnamefont
  {Lidar}},\ }\bibfield  {title} {\bibinfo {title} {Quantum-process tomography:
  {{Resource}} analysis of different strategies},\ }\href
  {https://doi.org/10.1103/PhysRevA.77.032322} {\bibfield  {journal} {\bibinfo
  {journal} {Physical Review A}\ }\textbf {\bibinfo {volume} {77}},\ \bibinfo
  {pages} {032322} (\bibinfo {year} {2008})}\BibitemShut {NoStop}%
\bibitem [{\citenamefont {Hu}\ \emph {et~al.}(2025)\citenamefont {Hu},
  \citenamefont {Zheng}, \citenamefont {Wang}, \citenamefont {Zhang},
  \citenamefont {Xu},\ and\ \citenamefont {Wang}}]{jkf7-wfcn}%
  \BibitemOpen
  \bibfield  {author} {\bibinfo {author} {\bibfnamefont {Y.}~\bibnamefont
  {Hu}}, \bibinfo {author} {\bibfnamefont {C.}~\bibnamefont {Zheng}}, \bibinfo
  {author} {\bibfnamefont {X.}~\bibnamefont {Wang}}, \bibinfo {author}
  {\bibfnamefont {Z.}~\bibnamefont {Zhang}}, \bibinfo {author} {\bibfnamefont
  {P.}~\bibnamefont {Xu}},\ and\ \bibinfo {author} {\bibfnamefont
  {K.}~\bibnamefont {Wang}},\ }\bibfield  {title} {\bibinfo {title} {Quantum
  process overlapping tomography: Theory and experiment},\ }\href
  {https://doi.org/10.1103/jkf7-wfcn} {\bibfield  {journal} {\bibinfo
  {journal} {Phys. Rev. Appl.}\ }\textbf {\bibinfo {volume} {23}},\ \bibinfo
  {pages} {064042} (\bibinfo {year} {2025})}\BibitemShut {NoStop}%
\bibitem [{\citenamefont {White}\ \emph {et~al.}(2020)\citenamefont {White},
  \citenamefont {Hill}, \citenamefont {Pollock}, \citenamefont {Hollenberg},\
  and\ \citenamefont {Modi}}]{white2020Demonstrationa}%
  \BibitemOpen
  \bibfield  {author} {\bibinfo {author} {\bibfnamefont {G.~A.}\ \bibnamefont
  {White}}, \bibinfo {author} {\bibfnamefont {C.~D.}\ \bibnamefont {Hill}},
  \bibinfo {author} {\bibfnamefont {F.~A.}\ \bibnamefont {Pollock}}, \bibinfo
  {author} {\bibfnamefont {L.~C.}\ \bibnamefont {Hollenberg}},\ and\ \bibinfo
  {author} {\bibfnamefont {K.}~\bibnamefont {Modi}},\ }\bibfield  {title}
  {\bibinfo {title} {Demonstration of non-{{Markovian}} process
  characterisation and control on a quantum processor},\ }\href
  {https://doi.org/10.1038/s41467-020-20113-3} {\bibfield  {journal} {\bibinfo
  {journal} {Nature Communications}\ }\textbf {\bibinfo {volume} {11}},\
  \bibinfo {pages} {6301} (\bibinfo {year} {2020})}\BibitemShut {NoStop}%
\bibitem [{\citenamefont {White}\ \emph {et~al.}(2021)\citenamefont {White},
  \citenamefont {Pollock}, \citenamefont {Hollenberg}, \citenamefont {Hill},\
  and\ \citenamefont {Modi}}]{white2021Manybody}%
  \BibitemOpen
  \bibfield  {author} {\bibinfo {author} {\bibfnamefont {G.~A.}\ \bibnamefont
  {White}}, \bibinfo {author} {\bibfnamefont {F.~A.}\ \bibnamefont {Pollock}},
  \bibinfo {author} {\bibfnamefont {L.~C.}\ \bibnamefont {Hollenberg}},
  \bibinfo {author} {\bibfnamefont {C.~D.}\ \bibnamefont {Hill}},\ and\
  \bibinfo {author} {\bibfnamefont {K.}~\bibnamefont {Modi}},\ }\bibfield
  {title} {\bibinfo {title} {From many-body to many-time physics},\ }\href@noop
  {} {\bibfield  {journal} {\bibinfo  {journal} {arXiv preprint
  arXiv:2107.13934}\ } (\bibinfo {year} {2021})},\ \Eprint
  {https://arxiv.org/abs/2107.13934} {arxiv:2107.13934} \BibitemShut {NoStop}%
\bibitem [{\citenamefont {White}\ \emph {et~al.}(5 27)\citenamefont {White},
  \citenamefont {Pollock}, \citenamefont {Hollenberg}, \citenamefont {Modi},\
  and\ \citenamefont {Hill}}]{whiteNonMarkovianQuantumProcess2022}%
  \BibitemOpen
  \bibfield  {author} {\bibinfo {author} {\bibfnamefont {G.}~\bibnamefont
  {White}}, \bibinfo {author} {\bibfnamefont {F.}~\bibnamefont {Pollock}},
  \bibinfo {author} {\bibfnamefont {L.}~\bibnamefont {Hollenberg}}, \bibinfo
  {author} {\bibfnamefont {K.}~\bibnamefont {Modi}},\ and\ \bibinfo {author}
  {\bibfnamefont {C.}~\bibnamefont {Hill}},\ }\bibfield  {title} {\bibinfo
  {title} {Non-{{Markovian Quantum Process Tomography}}},\ }\href
  {https://doi.org/10.1103/PRXQuantum.3.020344} {\bibfield  {journal} {\bibinfo
   {journal} {PRX Quantum}\ }\textbf {\bibinfo {volume} {3}},\ \bibinfo {pages}
  {020344} (\bibinfo {year} {2022-05-27})}\BibitemShut {NoStop}%
\bibitem [{\citenamefont {Pollock}\ \emph {et~al.}(1 25)\citenamefont
  {Pollock}, \citenamefont {Rodríguez-Rosario}, \citenamefont {Frauenheim},
  \citenamefont {Paternostro},\ and\ \citenamefont
  {Modi}}]{pollockNonMarkovianQuantumProcesses2018}%
  \BibitemOpen
  \bibfield  {author} {\bibinfo {author} {\bibfnamefont {F.~A.}\ \bibnamefont
  {Pollock}}, \bibinfo {author} {\bibfnamefont {C.}~\bibnamefont
  {Rodríguez-Rosario}}, \bibinfo {author} {\bibfnamefont {T.}~\bibnamefont
  {Frauenheim}}, \bibinfo {author} {\bibfnamefont {M.}~\bibnamefont
  {Paternostro}},\ and\ \bibinfo {author} {\bibfnamefont {K.}~\bibnamefont
  {Modi}},\ }\bibfield  {title} {\bibinfo {title} {Non-{{Markovian}} quantum
  processes: {{Complete}} framework and efficient characterization},\ }\href
  {https://doi.org/10.1103/PhysRevA.97.012127} {\bibfield  {journal} {\bibinfo
  {journal} {Physical Review A}\ }\textbf {\bibinfo {volume} {97}},\ \bibinfo
  {pages} {012127} (\bibinfo {year} {2018-01-25})}\BibitemShut {NoStop}%
\bibitem [{\citenamefont {Milz}\ \emph {et~al.}(2018)\citenamefont {Milz},
  \citenamefont {Pollock},\ and\ \citenamefont
  {Modi}}]{milz2018Reconstructing}%
  \BibitemOpen
  \bibfield  {author} {\bibinfo {author} {\bibfnamefont {S.}~\bibnamefont
  {Milz}}, \bibinfo {author} {\bibfnamefont {F.~A.}\ \bibnamefont {Pollock}},\
  and\ \bibinfo {author} {\bibfnamefont {K.}~\bibnamefont {Modi}},\ }\bibfield
  {title} {\bibinfo {title} {Reconstructing non-{{Markovian}} quantum dynamics
  with limited control},\ }\href {https://doi.org/10.1103/PhysRevA.98.012108}
  {\bibfield  {journal} {\bibinfo  {journal} {Phys. Rev. A}\ }\textbf {\bibinfo
  {volume} {98}},\ \bibinfo {pages} {012108} (\bibinfo {year}
  {2018})}\BibitemShut {NoStop}%
\bibitem [{\citenamefont {Li}\ \emph {et~al.}(2024)\citenamefont {Li},
  \citenamefont {Zheng}, \citenamefont {Meng}, \citenamefont {Zeng},
  \citenamefont {Luan}, \citenamefont {Zhang},\ and\ \citenamefont
  {Yu}}]{li2024non}%
  \BibitemOpen
  \bibfield  {author} {\bibinfo {author} {\bibfnamefont {Z.-T.}\ \bibnamefont
  {Li}}, \bibinfo {author} {\bibfnamefont {C.-C.}\ \bibnamefont {Zheng}},
  \bibinfo {author} {\bibfnamefont {F.-X.}\ \bibnamefont {Meng}}, \bibinfo
  {author} {\bibfnamefont {H.}~\bibnamefont {Zeng}}, \bibinfo {author}
  {\bibfnamefont {T.}~\bibnamefont {Luan}}, \bibinfo {author} {\bibfnamefont
  {Z.-C.}\ \bibnamefont {Zhang}},\ and\ \bibinfo {author} {\bibfnamefont
  {X.-T.}\ \bibnamefont {Yu}},\ }\bibfield  {title} {\bibinfo {title}
  {Non-markovian quantum gate set tomography},\ }\href@noop {} {\bibfield
  {journal} {\bibinfo  {journal} {Quantum Science and Technology}\ }\textbf
  {\bibinfo {volume} {9}},\ \bibinfo {pages} {035027} (\bibinfo {year}
  {2024})}\BibitemShut {NoStop}%
\bibitem [{\citenamefont {White}\ \emph {et~al.}(2023)\citenamefont {White},
  \citenamefont {Jurcevic}, \citenamefont {Hill},\ and\ \citenamefont
  {Modi}}]{white2023Unifying}%
  \BibitemOpen
  \bibfield  {author} {\bibinfo {author} {\bibfnamefont {G.~A.}\ \bibnamefont
  {White}}, \bibinfo {author} {\bibfnamefont {P.}~\bibnamefont {Jurcevic}},
  \bibinfo {author} {\bibfnamefont {C.~D.}\ \bibnamefont {Hill}},\ and\
  \bibinfo {author} {\bibfnamefont {K.}~\bibnamefont {Modi}},\ }\bibfield
  {title} {\bibinfo {title} {Unifying non-{{Markovian}} characterisation with
  an efficient and self-consistent framework},\ }\href@noop {} {\bibfield
  {journal} {\bibinfo  {journal} {arXiv preprint arXiv:2312.08454}\ } (\bibinfo
  {year} {2023})},\ \Eprint {https://arxiv.org/abs/2312.08454}
  {arxiv:2312.08454} \BibitemShut {NoStop}%
\bibitem [{\citenamefont {Greenbaum}(2015)}]{greenbaum2015Introduction}%
  \BibitemOpen
  \bibfield  {author} {\bibinfo {author} {\bibfnamefont {D.}~\bibnamefont
  {Greenbaum}},\ }\href@noop {} {\bibinfo {title} {Introduction to {{Quantum
  Gate Set Tomography}}}} (\bibinfo {year} {2015}),\ \Eprint
  {https://arxiv.org/abs/1509.02921} {arxiv:1509.02921 [quant-ph]} \BibitemShut
  {NoStop}%
\bibitem [{\citenamefont {Nielsen}\ \emph {et~al.}(2021)\citenamefont
  {Nielsen}, \citenamefont {Gamble}, \citenamefont {Rudinger}, \citenamefont
  {Scholten}, \citenamefont {Young},\ and\ \citenamefont
  {{Blume-Kohout}}}]{nielsen2021Gate}%
  \BibitemOpen
  \bibfield  {author} {\bibinfo {author} {\bibfnamefont {E.}~\bibnamefont
  {Nielsen}}, \bibinfo {author} {\bibfnamefont {J.~K.}\ \bibnamefont {Gamble}},
  \bibinfo {author} {\bibfnamefont {K.}~\bibnamefont {Rudinger}}, \bibinfo
  {author} {\bibfnamefont {T.}~\bibnamefont {Scholten}}, \bibinfo {author}
  {\bibfnamefont {K.}~\bibnamefont {Young}},\ and\ \bibinfo {author}
  {\bibfnamefont {R.}~\bibnamefont {{Blume-Kohout}}},\ }\bibfield  {title}
  {\bibinfo {title} {Gate {{Set Tomography}}},\ }\href
  {https://doi.org/10.22331/q-2021-10-05-557} {\bibfield  {journal} {\bibinfo
  {journal} {Quantum}\ }\textbf {\bibinfo {volume} {5}},\ \bibinfo {pages}
  {557} (\bibinfo {year} {2021})},\ \Eprint {https://arxiv.org/abs/2009.07301}
  {arxiv:2009.07301 [quant-ph]} \BibitemShut {NoStop}%
\bibitem [{\citenamefont {Viñas}\ and\ \citenamefont {Bermudez}(2
  06)}]{vinasMicroscopicParametrizationsGate2025}%
  \BibitemOpen
  \bibfield  {author} {\bibinfo {author} {\bibfnamefont {P.}~\bibnamefont
  {Viñas}}\ and\ \bibinfo {author} {\bibfnamefont {A.}~\bibnamefont
  {Bermudez}},\ }\bibfield  {title} {\bibinfo {title} {Microscopic
  parametrizations for gate set tomography under coloured noise},\ }\href
  {https://doi.org/10.1038/s41534-025-00976-4} {\bibfield  {journal} {\bibinfo
  {journal} {npj Quantum Information}\ }\textbf {\bibinfo {volume} {11}},\
  \bibinfo {pages} {23} (\bibinfo {year} {2025-02-06})}\BibitemShut {NoStop}%
\bibitem [{\citenamefont {Abrudan}\ \emph {et~al.}(2008)\citenamefont
  {Abrudan}, \citenamefont {Eriksson},\ and\ \citenamefont
  {Koivunen}}]{Abrudan2008}%
  \BibitemOpen
  \bibfield  {author} {\bibinfo {author} {\bibfnamefont {T.~E.}\ \bibnamefont
  {Abrudan}}, \bibinfo {author} {\bibfnamefont {J.}~\bibnamefont {Eriksson}},\
  and\ \bibinfo {author} {\bibfnamefont {V.}~\bibnamefont {Koivunen}},\
  }\bibfield  {title} {\bibinfo {title} {Steepest descent algorithms for
  optimization under unitary matrix constraint},\ }\href
  {https://doi.org/10.1109/TSP.2007.908999} {\bibfield  {journal} {\bibinfo
  {journal} {IEEE Transactions on Signal Processing}\ }\textbf {\bibinfo
  {volume} {56}},\ \bibinfo {pages} {1134} (\bibinfo {year}
  {2008})}\BibitemShut {NoStop}%
\bibitem [{\citenamefont {Wen}\ and\ \citenamefont {Yin}(2013)}]{Wen2013}%
  \BibitemOpen
  \bibfield  {author} {\bibinfo {author} {\bibfnamefont {Z.}~\bibnamefont
  {Wen}}\ and\ \bibinfo {author} {\bibfnamefont {W.}~\bibnamefont {Yin}},\
  }\bibfield  {title} {\bibinfo {title} {A feasible method for optimization
  with orthogonality constraints},\ }\href
  {https://doi.org/10.1007/s10107-012-0584-1} {\bibfield  {journal} {\bibinfo
  {journal} {Mathematical Programming}\ }\textbf {\bibinfo {volume} {142}},\
  \bibinfo {pages} {397} (\bibinfo {year} {2013})}\BibitemShut {NoStop}%
\bibitem [{\citenamefont {Zhu}\ \emph {et~al.}(2025)\citenamefont {Zhu},
  \citenamefont {Zhang}, \citenamefont {An},\ and\ \citenamefont
  {Zeng}}]{Zhu2025}%
  \BibitemOpen
  \bibfield  {author} {\bibinfo {author} {\bibfnamefont {X.}~\bibnamefont
  {Zhu}}, \bibinfo {author} {\bibfnamefont {C.}~\bibnamefont {Zhang}}, \bibinfo
  {author} {\bibfnamefont {Z.}~\bibnamefont {An}},\ and\ \bibinfo {author}
  {\bibfnamefont {B.}~\bibnamefont {Zeng}},\ }\bibfield  {title} {\bibinfo
  {title} {Unified framework for calculating convex roof resource measures},\
  }\href {https://doi.org/10.1038/s41534-025-01012-1} {\bibfield  {journal}
  {\bibinfo  {journal} {npj Quantum Information}\ }\textbf {\bibinfo {volume}
  {11}},\ \bibinfo {pages} {56} (\bibinfo {year} {2025})}\BibitemShut {NoStop}%
\bibitem [{\citenamefont {Ahmed}\ \emph {et~al.}(2023)\citenamefont {Ahmed},
  \citenamefont {Quijandr\'{\i}a},\ and\ \citenamefont
  {Kockum}}]{PhysRevLett.130.150402}%
  \BibitemOpen
  \bibfield  {author} {\bibinfo {author} {\bibfnamefont {S.}~\bibnamefont
  {Ahmed}}, \bibinfo {author} {\bibfnamefont {F.}~\bibnamefont
  {Quijandr\'{\i}a}},\ and\ \bibinfo {author} {\bibfnamefont {A.~F.}\
  \bibnamefont {Kockum}},\ }\bibfield  {title} {\bibinfo {title}
  {Gradient-descent quantum process tomography by learning kraus operators},\
  }\href {https://doi.org/10.1103/PhysRevLett.130.150402} {\bibfield  {journal}
  {\bibinfo  {journal} {Phys. Rev. Lett.}\ }\textbf {\bibinfo {volume} {130}},\
  \bibinfo {pages} {150402} (\bibinfo {year} {2023})}\BibitemShut {NoStop}%
\bibitem [{\citenamefont {Li}\ \emph {et~al.}(2025)\citenamefont {Li},
  \citenamefont {He}, \citenamefont {Zheng}, \citenamefont {Dong},
  \citenamefont {Luan}, \citenamefont {Yu},\ and\ \citenamefont
  {Zhang}}]{PhysRevLett.134.010803}%
  \BibitemOpen
  \bibfield  {author} {\bibinfo {author} {\bibfnamefont {Z.-T.}\ \bibnamefont
  {Li}}, \bibinfo {author} {\bibfnamefont {X.-L.}\ \bibnamefont {He}}, \bibinfo
  {author} {\bibfnamefont {C.-C.}\ \bibnamefont {Zheng}}, \bibinfo {author}
  {\bibfnamefont {Y.-Q.}\ \bibnamefont {Dong}}, \bibinfo {author}
  {\bibfnamefont {T.}~\bibnamefont {Luan}}, \bibinfo {author} {\bibfnamefont
  {X.-T.}\ \bibnamefont {Yu}},\ and\ \bibinfo {author} {\bibfnamefont {Z.-C.}\
  \bibnamefont {Zhang}},\ }\bibfield  {title} {\bibinfo {title} {Quantum comb
  tomography via learning isometries on stiefel manifold},\ }\href
  {https://doi.org/10.1103/PhysRevLett.134.010803} {\bibfield  {journal}
  {\bibinfo  {journal} {Phys. Rev. Lett.}\ }\textbf {\bibinfo {volume} {134}},\
  \bibinfo {pages} {010803} (\bibinfo {year} {2025})}\BibitemShut {NoStop}%
\bibitem [{\citenamefont {Boumal}(3
  16)}]{boumalIntroductionOptimizationSmooth2023a}%
  \BibitemOpen
  \bibfield  {author} {\bibinfo {author} {\bibfnamefont {N.}~\bibnamefont
  {Boumal}},\ }\href {https://doi.org/10.1017/9781009166164} {\emph {\bibinfo
  {title} {An {{Introduction}} to {{Optimization}} on {{Smooth Manifolds}}}}},\
  \bibinfo {edition} {1st}\ ed.\ (\bibinfo  {publisher} {Cambridge University
  Press},\ \bibinfo {year} {2023-03-16})\BibitemShut {NoStop}%
\bibitem [{\citenamefont {Reddi}\ \emph {et~al.}(2019)\citenamefont {Reddi},
  \citenamefont {Kale},\ and\ \citenamefont
  {Kumar}}]{reddi2019convergenceadam}%
  \BibitemOpen
  \bibfield  {author} {\bibinfo {author} {\bibfnamefont {S.~J.}\ \bibnamefont
  {Reddi}}, \bibinfo {author} {\bibfnamefont {S.}~\bibnamefont {Kale}},\ and\
  \bibinfo {author} {\bibfnamefont {S.}~\bibnamefont {Kumar}},\ }\href
  {https://arxiv.org/abs/1904.09237} {\bibinfo {title} {On the convergence of
  adam and beyond}} (\bibinfo {year} {2019}),\ \Eprint
  {https://arxiv.org/abs/1904.09237} {arXiv:1904.09237 [cs.LG]} \BibitemShut
  {NoStop}%
\bibitem [{\citenamefont {Gheorghiu}\ and\ \citenamefont
  {Lamata}(2018)}]{Gheorghiu2018}%
  \BibitemOpen
  \bibfield  {author} {\bibinfo {author} {\bibfnamefont {V.}~\bibnamefont
  {Gheorghiu}}\ and\ \bibinfo {author} {\bibfnamefont {L.}~\bibnamefont
  {Lamata}},\ }\bibfield  {title} {\bibinfo {title} {Quantum++: A modern c++
  quantum computing library},\ }\href
  {https://doi.org/10.1371/journal.pone.0208073} {\bibfield  {journal}
  {\bibinfo  {journal} {PLoS ONE}\ }\textbf {\bibinfo {volume} {13}},\ \bibinfo
  {pages} {e0208073} (\bibinfo {year} {2018})}\BibitemShut {NoStop}%
\end{thebibliography}%

\appendix
\section{Proof of Theorem 1}\label{app1}
The proof proceeds by showing that each component of the CIS set 
$\Gamma=\{\mathcal{C}^{ (N)},\mathfrak{J},\mathfrak{S}\}$ admits a 
Stiefel-manifold parametrization, and that the product of these 
manifolds yields the geometry in~\eqref{eq:cis_prod_st}.

\subsection{Parameterization of the comb.}
For a finite-memory $N$-slot quantum comb $\mathcal{C}^{ (N)}$ with  the causal structure, it can be completely 
characterized by the concatenation of $N$ isometries ~\cite{Bisio2011}.
Thus, for every slot $t$, there exists an 
isometry
$V^{ (t)}:\mathcal{H}_{\tti_t}\otimes \mathcal{H}_{a_t} \rightarrow \mathcal{H}_{\tto_t}\otimes \mathcal{H}_{a_{t+1}}$
such that the entire comb can be written as
\begin{equation}
\begin{aligned}
\mathcal{C}^{ (N)} (\rho_{\tti}) 
= \Tr_{a_{N-1}}\!\Bigl[
V^{ (N-1)}\cdots V^{ (0)} \,
\rho_{\tti}\,
V^{ (0)\dagger}\cdots V^{ (N-1)\dagger}
\Bigr].
\end{aligned}
\label{eq:comb_iso_rep_proof}
\end{equation}
  
Because $V^{ (t)}$ is an isometry satisfying $V^{ (t)\dagger}V^{ (t)}=\idop_{d_{\tti_{t}}d_{a_t}}$, we have $
V^{ (t)}\in \mathrm{St} (d_{\tti_t} d_{a_t},\, d_{\tto_t} d_{a_{t+1}})$.
Therefore, the entire sequence 
$\mathbb{V}= (V^{ (0)},\dots,V^{ (N-1)})$ lies on the product manifold  
\[
\bigtimes_{t=0}^{N-1}\mathrm{St} (d_{\tti_{t}}d_{{a_t}},\, d_{\tto_{t}}d_{{a_{t+1}}}).
\]

Theorem 2 in ~\cite{Bisio2011} shows that the minimal dimension of the ancilla space $\cH _{a_{t+1}}$ is the dimension of the support of $\Upsilon^{ (t+1)}$.
So the ancilla dimensions satisfy 
$d_{a_{t+1}}\ge \rank (\Upsilon^{ (t+1)})$, ensuring that each comb link admits such an isometry.

\subsection{Parameterization of instruments.}
Consider an instrument $\mathcal{J}^{ (t)}=\{\mathcal{A}^{ (t)}_{x}\}$ 
at time step $t$.  
Each branch $\mathcal{A}^{ (t)}_{x}$ is a CPTNI map.
It admits a Stinespring dilation
\begin{equation}
    \mathcal{A}^{ (t)}_{x} (\rho)=
\Tr_{e_{t,x}}
\bigl[W^{ (t)}_{x}\,\rho\,W^{ (t)\dagger}_{x}\bigr],
\end{equation}
where $\cH_{e_{t, x}}$ is the ancillary dimension of the instrument, and $W^{ (t)}_{v,x}:\mathcal{H}_{\tto_t}\rightarrow \mathcal{H}_{\tti_{t+1}} \otimes \mathcal{H}_{e_{t,v,x}}$.
The completeness of the instrument,$\sum_x \mathcal{A}^{ (t)}_{x} $  is trace preserving, implies the condition
$\sum_x W^{ (t)\dagger}_{x} W^{ (t)}_{x}=\idop_{d_{\tto_t}}$. 

Note that for any input state $\rho \in \Lin (\cH_{\tto_t})$, 
the output $\cA^{ (t)}_{x} (\rho)$ is a positive semidefinite operator 
on $\cH_{\tti_{t+1}}$ with trace
\begin{equation}
    p (x|\rho) \;=\; 
    \Tr\!\left[\cA^{ (t)}_{x} (\rho)\right],
\end{equation}
which has the operational meaning of the probability of observing an outcome.
Thus $\cA^{ (t)}_{x} (\rho)$ encodes both the probability of the classical 
outcome $x$ and the corresponding  (unnormalized) post-measurement quantum 
state. The normalized state conditioned on obtaining outcome $x$ is given by
\begin{equation}
    \rho_{x}
    =\frac{\cA^{ (t)}_{x} (\rho)}{p (x|\rho)}.
\end{equation}
This explicitly shows that each $\cA^{ (t)}_{x}$ is CPTNI, and the family 
$\{\cA^{ (t)}_{x}\}$ forms a quantum instrument.

In order to embed these variables onto the manifold, stacking all branches yields the block operator
\begin{equation}
J^{ (t)} := [W^{ (t)}_{1},\ldots, W^{ (t)}_{|\mathcal{J}^{ (t)}|}],   
\end{equation}
which satisfies $J^{ (t)\dagger}J^{ (t)}=\idop_{d_{\tto_t}}$.  
Thus, the quantum instrument $J^{ (t)}$ is on the $\mathrm{St}\!\left (\sum_{x=1}^{|\mathcal{J}^{ (t)}|} d_{\tti_{t+1}}d_{e_{t,x}},\; d_{\tto_t}\right)$.

Then consider the quantum instrument set $\fJ^{ (t)}=\{\cJ^{ (t)}_v\}$ at each slot from $t=0$ to $t=N-1$, the collection of all there instrument is given as $\bJ= (\fJ^{ (0)},...,\fJ^{ (N-1)})$. It can be parameterized on the manifold
\[
\bigtimes_{t=0}^{N-1}\bigtimes_{v=1}^{|\mathfrak{J}^{ (t)}|}
\mathrm{St}\!\left (
\sum_{x=1}^{|\mathcal{J}^{ (t)}_{v}|} d_{\tti_{t+1}}d_{e_{t,v,x}}, 
\; d_{\tto_t}
\right),
\]
where dimensions satisfying $d_{e_{t,v,x}} \ge \rank (A^{ (t)}_{v,x})$.

\subsection{Parameterization of initial states.}
Each possibly mixed initial state 
$\rho^{ (0)}_u\in\Lin (\mathcal{H}_{\tti_0})$ admits a purification
\begin{equation}
    \rho^{ (0)}_u = \Tr_{r_u} (S_u S_u^\dagger),
\end{equation}
where $S_u$ is the pure state in the composite system $\mathcal{H}_{\tti_0}\otimes \mathcal{H}_{r_u}$,
and dimension $d_{r_u}\ge\rank (\rho^{ (0)}_u)$.

There is the constraint $
S_u^\dagger S_u =1$.
Hence,
$S_u$ is on the $\mathrm{St} (d_{\tti_0} d_{r_u},1)$. The product manifold of the state set $\bS= (S_1, ..., S_{|\fS|})$ is 
\[
\bigtimes_{u=1}^{|\mathfrak{S}|} 
\mathrm{St} (d_{\tti_0}d_{r_u},1).
\]

Since the comb, instrument, and initial-state components are mutually independent degrees of freedom, every admissible CIS set $\Gamma$ corresponds to a unique point on this product Stiefel manifold, and conversely every point on this manifold generates a valid CIS set.

\section{Riemannian gradient on Stiefel manifold}
Since the instrument set is on the Stiefel manifold, $\mathcal L (\mathfrak{A}_\mathrm{St})$ can be optimized via unconstrained Stiefel manifold optimization with natural satisfaction of physical constraints. In this work, we adopt the adaptive moment estimation  (ADAM) algorithm adapted to the geometry of the Stiefel manifold. While ADAM is traditionally formulated for Euclidean spaces, it can be generalized to Riemannian manifolds by projecting gradients onto tangent spaces and updating parameters via suitable retractions.

For Riemannian submanifolds, the Riemannian gradient is the orthogonal projection of the “classical” gradient to the tangent spaces ~\cite{boumalIntroductionOptimizationSmooth2023a}.
Let $\mathrm{St} (n, p)$ denote the Stiefel manifold, i.e., the set of all $n \times p$ real matrices with orthonormal columns. The \textit{tangent space} at a point $X \in \mathrm{St} (n,p)$ is given by:
\begin{align}
  \cT_X\mathrm{St} (n,p)=
\begin{Bmatrix}
V\in\mathbb{C}^{n\times p}:X^\dagger V+V^\dagger X=0
\end{Bmatrix}.
\end{align}

The orthogonal projector to a tangent space of $\mathrm{St} (n,p)$ must be such that $U-\mathrm{Proj}_X (U)$  is orthogonal to $\cT_X\mathrm{St} (n,p)$, that is, the difference must be in the orthogonal complement of the tangent space. For a Euclidean gradient $G \in \mathbb{C}^{n \times p}$ of cost function, its \textit{Riemannian gradient} on the Stiefel manifold is obtained by projecting $G$ orthogonally onto the tangent space:
\begin{align}
\nabla^{\mathrm{St}}_X \mathcal{L} = \mathrm{Proj}_X (G)= G - X \mathrm{sym} (X^\dagger G),
\end{align}
where $\mathrm{sym} (X) := \frac{1}{2} (X + X^\dagger)$. This projection ensures that updates remain tangent to the manifold.

To move along the manifold after computing a gradient step, we need a retraction operation. Among several retraction choices  (e.g., polar decomposition, QR factorization), we use the Cayley transform due to its efficiency and smoothness. Given a step size $\tau$ and a skew-symmetric matrix $D \in \mathbb{C}^{n \times p}$, the Cayley retraction is defined as:
\begin{align}
  X' =  (I + \frac{\tau}{2} D)^{-1}  (I - \frac{\tau}{2} D) X,
\end{align}
where setting $D = \nabla^{\mathrm{St}}_X \mathcal{L} X^\dagger - X  (\nabla^{\mathrm{St}}_X \mathcal{L})^\dagger$ enables updates for variables on the Riemannian manifold. These fundamental principles allow direct application of various optimization algorithms.

\section{Stiefel ADAM algorithm}
The ADAM algorithm serves as an effective alternative to stochastic gradient descent  (SGD) in classical optimization problems ~\cite{reddi2019convergenceadam}. By incorporating momentum methods, it dynamically adapts learning rates during the training process. At each iteration, the optimizer conducts the following procedures with hyperparameters $\gamma_1$, $\gamma_2$, $\tau_0$, and $\epsilon$:

\begin{algorithm}[h]
  \caption{ADAM on the Stiefel manifold}\label{alg:adam_stiefel}
  Initialize parameter and first and second moments;\\
  Compute the Euclidean gradient $\nabla_{\Gamma} \mathcal{F}$ of cost function $\mathcal{F}$;\\
  \While{not converged}{
    Update: $M_G \leftarrow \gamma_1 M_G + { (1-\gamma_1)\nabla_{\Gamma} \mathcal{F}}$;\\    
    Update:  $v\leftarrow \gamma_2 v+  (1-\gamma_2) \|\nabla_{\Gamma} \mathcal{F}\|^2_F$;\\
    Estimate biased-corrected ratio:  $r\leftarrow (1-\gamma_1^t)\sqrt{v_t/ (1-\gamma_2^t)+\epsilon}$;\\
    Project onto the tangent space:  $G_\mathrm{St} \leftarrow \frac{1}{r} (M_G -\Gamma ~ \mathrm{sym} (\Gamma ^\dagger M_G))$;\\ 
    Select adaptive learning rate:  $\tau\leftarrow\min\{\tau_0,1/ (\|G_\mathrm{St}\|_F+\epsilon)\}$;\\
    Cayley retraction: $\Gamma \leftarrow  (I+\tau\frac{D}{2})^{-1} (I-\tau\frac{D}{2})\Gamma$;\tcp*[f]{$D$ corresponding to $G_\mathrm{St}$ }
  }
  \Return{$\Gamma$.}
\end{algorithm}

Note that the initialization of variables in the algorithm can be performed either randomly or based on prior experimental knowledge. When no prior information is available about the instruments, one can first apply a Linear Inversion method to obtain a rough estimate and then project it onto the Stiefel manifold. This initialization strategy helps accelerate convergence during subsequent manifold-based optimization.

Since all variables reside on a common manifold, they can be optimized jointly at each iteration, facilitating convergence toward a globally consistent solution. While this work uses ADAM due to its stability and efficiency, other Riemannian optimization methods, such as Riemannian conjugate gradient or trust-region approaches, can also be considered depending on problem complexity and convergence requirements.

\section{Complexity analysis}
We now analyze the computational complexity of the proposed framework. The primary computational cost arises from gradient evaluations during each iteration of the optimization process. In our method, multiple variables lie on the Stiefel manifold, and thus we analyze the gradient expressions for each variable individually. The variable with the most complex gradient expression provides an upper bound for the overall per-iteration complexity. 

The gradient with respect to the measurement matrix $\mathbb{J}_x^{ (t)}$ involves a double summation over time and CPTNI maps indices. It can be seen as two parts: the CPTP map and the POVM. We analyze the latter part first, $M_{v,x}$ denotes the POVM in the $\cA_{v,x}$. The Euclidean gradient of $M_{v,x}$ is
\begin{widetext}

\begin{align}
    G_{M_{v,x}} = \nabla_{M_{v,x}^{ (t)}}\mathcal{L}^{ (t)} = \sum_{\bm{x}} 4p_{\bm{x}} ^{ (t-1)}\left (p_{\bm{x}} ^{ (t-1)}\Tr[M^{ (t)}_{v,x}\eta^{ (t)}_{\mathtt{i},\bm{x}}M^{ (t)\dagger}_{v,x}] -\tilde{p}^{ (t)}_{\bm{x}}\right)M^{ (t)}_{v,x}\eta^{ (t)}_{\mathtt{i},\bm{x}}. 
\end{align}
    
\end{widetext}
then $\mathbb{G}_{M_x} =[G_{M_{x_1}}, \dots, G_{M_{x_m}}]$ corresponding to the cost function. 

In addition, the gradient with respect to the CPTP part is computed by differentiating over each element in the set $\{U_v\}$ :
\begin{widetext}
    
\begin{align}
    \nabla_{ U^{ (t-1)}} \mathcal{L}^{ (t)}=\sum_{\bm{x}} \sum_{v}\sum_{x} 4p_{{\bm{x}}} ^{ (t-1)}\left (p_{{\bm{x}}} ^{ (t-1)}\Tr[W^{ (t)}_{v,x}U^{ (t-1)}_{v}\eta^{ (t-1)}_{\mathtt{i},\bm{x}} U^{ (t-1)\dagger}_{v} ]\right.\notag -  \left. \tilde{p}^{ (t)}_{\bm{x}}\right)W^{ (t)}_{v,x} U_{v}^{ (t-1)} \eta^{ (t-1)}_{\mathtt{i},\bm{x}},
\end{align}
\end{widetext}
\text{where} $W^{ (t)}_{v,x} =V^{ (t)\dagger}M_{v,x}^{ (t)\dagger} M_{v,x}^{ (t)}V^{ (t)}$. 

To estimate the overall complexity, assume the input and output dimensions are all $d$. Let $T_t$ denote the number of iterations required for convergence at time step $t$, and $n_A$ the number of available instruments  (with $n_m < n_A$ denoting the number of measurements). Then, the total complexity of estimating the full instrument set is given by Eq.~\eqref{complexity}.
This estimate accounts for the cost of matrix multiplications and trace evaluations at each gradient step. The first term corresponds to the complexity of updating variables, while the second term captures the cost associated with measurement gradients.

Compared to maximum likelihood estimation-based frameworks—which require constrained optimization over high-dimensional parameter spaces—our method is significantly more scalable. MLE approaches typically incur exponential complexity due to the explicit enforcement of physical constraints.

Moreover, our method enables further complexity reduction through dimensional compression of the ancillary Hilbert spaces in the quantum comb, informed by prior knowledge. This ability to adaptively optimize ancillary dimensions underscores the intrinsic scalability advantage of our Riemannian geometry-based framework.

\end{document}